\documentclass[
floatfix,twocolumn,superscriptaddress, pra, aps
 amsmath,amssymb,
 aps,
]{revtex4-2}
\usepackage[utf8]{inputenc}
\usepackage{amsmath}
\usepackage{amsfonts}
\usepackage{amssymb,amsthm}
\usepackage[subrefformat=parens,labelformat=parens,caption=false]{subfig}
\usepackage{graphicx}
\usepackage{multirow}
\usepackage[]{hyperref}
\usepackage{bbold}
\usepackage{cancel}

\hypersetup{
  colorlinks = true,
  urlcolor = magenta,
  linkcolor = red,
  citecolor = blue
}
\usepackage[normalem]{ulem}
\usepackage{braket}
\usepackage[all]{hypcap}
\usepackage{url}
\usepackage{float}

\usepackage{xcolor}

\usepackage[]{hyperref}
\usepackage{bbold}
\usepackage{bm}
\usepackage{cancel}
\newcommand{\eq}[1]{\begin{align}#1\end{align}}
\newcommand{\ie}{i.\,e.,\ }

\usepackage{graphicx}
\graphicspath {{figures/}}

\begin{document}
\title{Multi-excitation scattering in subwavelength atomic arrays
}

\author{Yidan Wang}
\affiliation{Department of Physics, Harvard University, Cambridge, Massachusetts 02138, USA}
\author{Oriol Rubies-Bigorda}
\affiliation{Physics Department, Massachusetts Institute of Technology, Cambridge, Massachusetts 02139, USA}

\author{Valentin Walther}
 \affiliation{Department of Physics and Astronomy, Purdue University, West Lafayette, Indiana 47907, USA}
 \affiliation{Department of Chemistry, Purdue University, West Lafayette, Indiana 47907, USA}

\author{Susanne Yelin}
\affiliation{Department of Physics, Harvard University, Cambridge, Massachusetts 02138, USA}

\begin{abstract}

Subwavelength atomic arrays are a leading platform for engineering light-matter interactions, enabling near-perfect single-photon mirrors and robust quantum memories based on long-lived dark spin waves. However, a comprehensive theory of their nonlinear, multi-excitation dynamics has remained a significant challenge. 
We present a unified quantum scattering theory that treats both photons and collective atomic spin waves as distinct propagating excitations interacting across different spatial dimensions. 
Our central result is a powerful analytical reduction: we demonstrate that the complete multi-channel S-matrix and the associated scattering cross sections are exactly determined by the effective scattering dynamics solely within the atomic spin wave subspace. This maps the complex physical problem of photon-atom interactions to a conceptually simpler one involving only atomic modes. We apply this formalism to the two-excitation case, deriving the complete analytical S-matrix and scattering cross sections for systems with two-level nonlinearities. Our work provides a versatile analytic tool for analyzing and engineering complex quantum nonlinear phenomena, including multi-excitation subradiance, in large-scale atomic systems.
\end{abstract}
\maketitle

\section{Introduction}

Subwavelength atomic arrays have emerged as a powerful platform for engineering strong light-matter interactions through quantum  interference.
In these systems, atoms positioned at subwavelength distances interact collectively via photons in free space, giving rise to collective atomic excitations with superradiant (enhanced) or subradiant (suppressed) decay rates \cite{garcia2007colloquium, shahmoon2017cooperative, bettles2016enhanced}. This collective behavior underpins two foundational phenomena: the ability to act as a perfect ``quantum mirror" that reflects single photons without loss into unwanted modes \cite{shahmoon2017cooperative,bettles2016enhanced, rui2020subradiant, srakaew2023subwavelength}, and the capacity to store quantum information in extremely long-lived ``dark spin waves", making them  useful for quantum memory applications \cite{ExponentialAsenjoGarcia2017, Manzoni2018}.  Fueled by recent experimental progress, harnessing these collective effects has pushed atomic arrays to the forefront of quantum optics, driving advances in nonlinear quantum optics, quantum information processing, and many-body physics \cite{bekenstein2020quantum, wei2021generation, Shah2024, Patti2021, moreno2021quantum, zhang2022photon, Rubies-Bigorda2025Deterministic,jaworowski2025laughlin}.

 While these phenomena are well-understood in the single-excitation limit \cite{zoubi2011lifetime, jenkins2012controlled, plankensteiner2015selective, bettles2015cooperative, bettles2016eigenmodes, bettles2016enhanced, sutherland2016collective}, a comprehensive theory for the multi-excitation regime presents a significant challenge. Current theoretical approaches are often constrained, relying on numerical methods limited by system size \cite{cidrim2020photon, bettles2020quantum, williamson2020superatom, moreno2021quantum, Robicheaux2023}, approximate methods \cite{moreno2021quantum, Ostermann2024, RubiesBigorda2023, Scarlatella2024, scarlatella2024subwavelength}, or analytical treatments restricted to specific scenarios. For instance, some analytical results are confined to 1D arrays \cite{Zhang2019, Iversen2021, ExponentialAsenjoGarcia2017}, while recent scattering formalisms for photons do not address atomic subradiance  \cite{Solomons2023, Pedersen2024}.

\begin{figure}
    \centering
\includegraphics[width=1\linewidth]{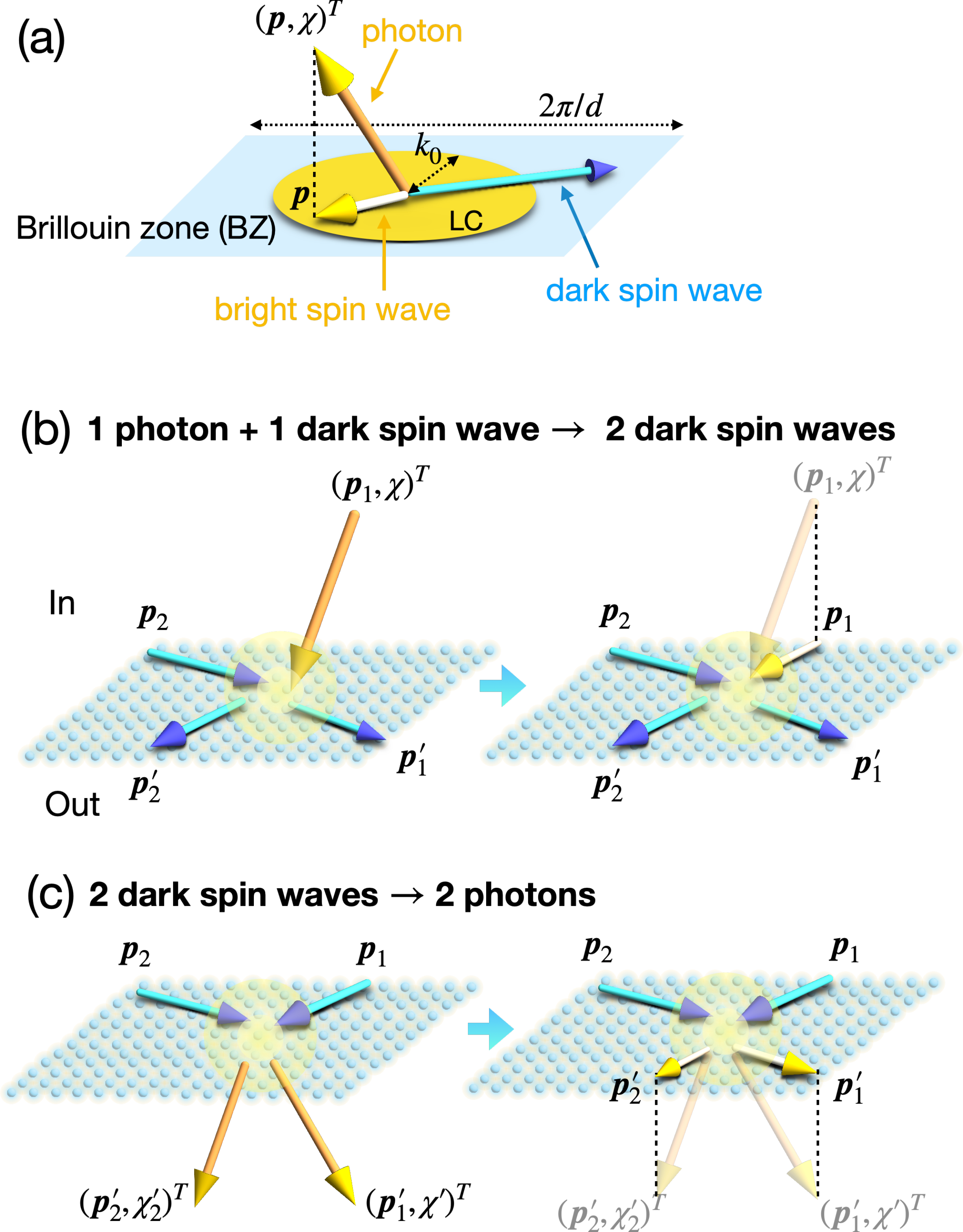}
    \caption{ \textbf{Scattering between photons and dark spin waves in 2D atomic arrays}
\textbf{(a)} Schematic illustration of spin waves and photons in momentum space. Dark and bright spin waves have 2D momenta within the Brillouin zone, lying outside and inside the light cone (LC, defined by $\bm{p} < k_0$, where $k_0$ is the wave number resonant with the atom transition), respectively. The momentum of a photon $(\bm{p},\chi)^T$ near atomic resonance is a 3D vector with out-of-plane component $\chi$ and  in-plane projection $\bm{p}$ within the light cone, corresponding to a bright spin wave. 
\textbf{(b,c)} Conceptual illustration of the scattering formalism: scattering processes involving photons (left) can be mapped to spin wave scattering (right) by replacing each photon with a bright spin wave of the same lattice momentum. 
\textbf{(b)} An incoming photon with momentum $(\bm{p}_1,\chi)^T$ scatters with a dark spin wave ($\bm{p}_2$), generating a pair of outgoing dark spin waves ($\bm{p}_1, \bm{p}_2$). 
\textbf{(c)} Two incoming dark spin waves with momenta $\bm{p}_1$ and $\bm{p}_2$ scatter into two photons with momenta $(\bm{p}'_1,\chi'_1)^T$ and $(\bm{p}'_2,\chi'_2)^T$. 
Both processes are computed entirely within the atomic subspace by replacing each photon with a bright spin wave of the same lattice momentum.
}
    \label{fig:scatteringIllu}
\end{figure}

In our work, we develop a scattering theory framework that captures nonlinear phenomena involving both multiple photons and subradiant spin waves as active participants in the scattering process, as we illustrate in Fig.\ \ref{fig:scatteringIllu}  for the case of 2D atomic arrays. In conventional light-matter interaction systems, local emitters are typically viewed as scatterers, while photons act as the propagating excitations that are scattered.  
However, a distinguishing feature of subwavelength atomic arrays is that the atoms are not merely passive scatterers. They also support long-lived, collective spin waves that act as propagating excitations themselves. In the thermodynamic limit considered here, these subradiant modes become perfectly stable, forming ``dark spin waves" that are completely decoupled from the radiative continuum. The ability of these dark spin waves to undergo scattering alongside photons leads to a mixed-dimensional multi-excitation problem, where photons propagate in 3D space while the spin waves are confined to the array.

Our theoretical framework is designed to capture the full range of these interactions. The scattering process can result in various combinations of outgoing photons and dark spin waves, and we formalize each of these possibilities as a distinct scattering channel. The main result of this work is a powerful analytical reduction of this complex multi-channel problem. We derive the two fundamental quantities of scattering theory: the S-matrix (Section  \ref{sec_scatt_matrix}), which encodes the transition amplitudes and phases between states in all channels, and the scattering cross sections (Section \ref{sec_cs}), which quantify the effective area of scattering for an incoming flux of excitations. Crucially, we show that both are fully determined by the scattering T-matrix defined purely within the atomic spin-wave subspace. To demonstrate the power of the formalism, we apply it to the important case of two-excitation scattering (Section \ref{subsectionTwoexci}). By focusing on systems with two-level nonlinearities, we derive a complete analytical expression for the S-matrix and scattering cross sections. Our formalism also establishes a clear framework for studying multi-excitation dark states.

Our framework is naturally applicable to infinite lattices, providing an analytical counterpart to numerical studies restricted to finite systems. It applies to multiple excitations in both 1D and 2D arrays with general atomic nonlinearities and is readily extendable to diverse level structures, lattice geometries, and sublattices. For instance, in a companion paper \cite{Wang2025UniversalScattering}, we employ this formalism to analyze two-excitation scattering in 2D arrays, revealing the emergence of universal scattering features.

\section{System and single-excitation physics\label{secIntro}}
This section lays the foundation for our analysis by introducing the system Hamiltonian, reviewing single-excitation scattering, and presenting the effective spin model that serves as the fundamental tool for studying multi-excitation scattering.

 We consider an array of $N$ identically polarized two-level atoms with transition frequency $\omega_{eg}$, coupled to free-space photons. The two-level nature of each atom imposes a local hard-core constraint: no site can be occupied by more than one excitation. In the operator formalism, this means the atomic excitations are treated as hard-core bosons. Our analysis focuses on 1D and 2D square arrays with a sub-half-wavelength spacing $d < \lambda_0/2$ (where $\lambda_0 = 2\pi c/\omega_{eg}$), though the formalism can be readily extended to other geometries. In the thermodynamic limit ($N \to \infty$), the system's translational symmetry enables a description in terms of collective momentum eigenstates.

The full Hamiltonian for this system can be expressed as the sum of a quadratic part, describing linear dynamics, and a quartic part, accounting for nonlinear interactions: 
\begin{equation}
H = \int_{\mathrm{BZ}} d\bm{p}\ H_{\bm{p}} + U.\label{eqHamilSimp} 
\end{equation}
 The microscopic origins and formal definition of this Hamiltonian are detailed in Appendix~\ref{app:system_details}.
The nonlinear interaction term $U$ describes momentum-conserving scattering processes between atomic excitations, and is given by
\eq{
U = \frac{1}{2}\int d\bm{q} \, d\bm{q}'  & \int d\bm{P} \,   U(\bm{q} - \bm{q}')  \nonumber \\ 
& \times b^\dagger_{\boldsymbol{P}/2+\bm{q}'} \, b^\dagger_{\boldsymbol{P}/2-\bm{q}'} \, b_{\boldsymbol{P}/2-\bm{q}} b_{\boldsymbol{P}/2+\bm{q}}, 
\label{eq:U_interaction_intro}
}
where $b^\dagger_{\bm{p}}$ creates a collective atomic spin wave with lattice momentum $\bm{p}$ and satisfies the bosonic commutation relation $[b_{\bm{p}}, b^\dagger_{\boldsymbol{p}'}] = \delta(\bm{p} - \bm{p}')$.
The Brillouin zone (BZ) is the line segment $p \in [-\pi/d, \pi/d]$ for a 1D array and the square $p_x, p_y \in [-\pi/d, \pi/d]$ for a 2D array.

The hard-core constraint imposed by the two-level atoms corresponds to an infinite-strength, zero-range contact interaction. In the momentum representation, this simplifies to a momentum-independent potential, $U(\bm{q}) \to \infty$. While our analysis primarily focuses on this case, the formalism we develop applies to momentum-dependent interactions.\footnote{Let $\tilde{U}(\bm{r})$ be the real-space interaction potential corresponding to the momentum-space interaction $U(\bm{q})$. Our framework can accommodate other physically relevant potentials such as the van der Waals interaction, $\tilde{U}(\bm{r}) \propto 1/|\bm{r}|^6$, for Rydberg atoms, or the dipole-dipole interaction, $\tilde{U}(\bm{r}) \propto 1/|\bm{r}|^3$, for atoms with permanent dipole moments.}

 \subsection{Single-excitation scattering \label{sub_sec_single_excitation}}
 
The single-excitation (linear) dynamics, governed by $H_{\bm{p}}$, are determined by the coupling of atomic excitations to the continuum of photonic modes. This coupling crucially depends on whether the momentum $\bm{p}$ lies inside or outside the light cone (LC), defined by $|\bm{p}| = k_0$, where $k_0 = \omega_{eg}/c$ [see Fig.\ref{fig:scatteringIllu} (a)]. We work in a rotating frame at the atomic resonance frequency $\omega_{eg}$, effectively setting it as the zero of energy for excitations.

For momenta outside the light cone ($\bm{p} \notin \text{LC}$), atomic excitations are off-resonant from all propagating photon modes and are effectively decoupled from them. These modes, known as \emph{dark spin waves}, evolve under a simple Hamiltonian with real-energy dispersion
\begin{equation}
H_{\bm{p}} =  \Delta(\bm{p}) b^\dagger_{\bm{p}}b_{\bm{p}} \qquad (\bm{p} \notin \text{LC}),
\end{equation}
 where $\Delta(\bm{p})$ is the collective Lamb shift resulting from the virtual exchange of off-resonant photons. 

For momenta inside the light cone ($\bm{p} \in \text{LC}$), the atomic excitations, known as \emph{bright spin waves}, are resonantly coupled to propagating photon modes. By redefining the photonic mode basis, this interaction can be recast as a waveguide-QED-type model in which each bright spin wave $b^\dagger_{\bm{p}}$ couples to an effective 1D continuum of photons created by $C_{\bm{p}}^\dagger(\chi)$. These operators satisfy $[C_{\bm{p}}(\chi), C^\dagger_{\bm{p}'}(\chi')] = \delta(\bm{p}-\bm{p}')\delta(\chi-\chi')$.

For a 2D array in the $xy$-plane, $C^\dagger_{\bm{p}}(\chi)$ represents an equal superposition of photons propagating away from the array in the $+z$ and $-z$ directions, that is, with 3D wavevectors $(\bm{p}, \chi)$ and $(\bm{p}, -\chi)$, and a specific superposition of polarizations. For a 1D array aligned along the $z$-axis, $C^\dagger_{p}(\chi)$ describes a superposition of photons propagating in the azimuthal direction, including both polarizations. In both cases, $\chi > 0$ denotes the magnitude of the transverse wavenumber. Explicit expressions for $C^\dagger_{\bm{p}}(\chi)$ are provided in Appendix~\ref{app:system_details}.

For both 1D and 2D arrays, the Hamiltonian that governs the single-excitation dynamics for a given $\bm{p} \in \text{LC}$ is
\eq{
H_{\bm{p}} &= \int_{0}^{+\infty} d\chi \, E_{\bm{p}}(\chi) C_{\bm{p}}^{\dagger} (\chi) C_{\bm{p}}(\chi) +  \Delta(\bm{p}) b^\dagger_{\bm{p}}b_{\bm{p}} \nonumber \\
&\quad + \int_{0}^{+\infty} d\chi\, \big[ g_{\bm{p}}\, C^\dagger_{\bm{p}}(\chi) b_{\bm{p}} + \text{h.c.} \big]  \qquad (\bm{p} \in \text{LC}),
\label{eq:H_p_LC_intro}
}
where $E_{\bm{p}}(\chi) = c\sqrt{\lVert\bm{p}\rVert^2 + \chi^2}-\omega_{eg}$ is the photon dispersion relation in the rotating frame.

An important parameter characterizing the atom-photon coupling is $\Gamma(\bm{p})$, which denotes the decay rate of the bright spin-wave mode $b^\dagger_{\bm{p}}$ after tracing out all photon modes (see Refs.~\cite{Asenjo-Garcia2017, ExponentialAsenjoGarcia2017}). The coupling coefficient $g_{\bm{p}}$ in Eq.~\eqref{eq:H_p_LC_intro} is related to $\Gamma(\bm{p})$ via $g_{\bm{p}} = \sqrt{\Gamma(\bm{p}) v_{g} / (2\pi)}$, where $v_{g} = \partial_\chi E_{\bm{p}}(\chi) = \chi / \omega_{eg}$ is the photon group velocity.

For each incoming photon in the effective mode $C^\dagger_{\bm{p}}(\chi)$, there exists a corresponding scattering eigenstate of $H_{\bm{p}}$ with creation operator $\psi^\dagger_{\bm{p}}(\chi)$, which we refer to as a \emph{dressed-photon state}. This eigenstate contains an atomic component, which takes the form of a bright spin wave with amplitude $a_{\bm{p}}(E)$,
\begin{equation}
a_{\bm{p}}(E)= \frac{g_{\bm{p}}}{E-\epsilon(\bm{p})}.    \label{eqepchi} 
\end{equation}
Here, $E=E_{\bm{p}}(\chi)$ is the energy of the photon and $\epsilon(\bm{p}) \equiv \Delta(\bm{p}) -  i\Gamma(\bm{p})/2$ is the complex spin-wave dispersion [see Eq.~\eqref{eqepsilon}].
The dressed-photon state also contains a photonic component; the outgoing part of this photonic wavefunction acquires a phase shift relative to its incoming part, characterized by the transmission coefficient $t_{\bm{p}}(E)$
\begin{equation}
t_{\bm{p}}(E)=\frac{E-\epsilon(\bm{p})^*}{E -\epsilon(\bm{p})}. \label{eqt_pchi}
\end{equation}
This effective ``single-emitter'' Hamiltonian, in which the collective mode $b^\dagger_{\bm{p}}$ acts as the emitter, is Hermitian and gapless within the relevant energy window, and thus does not support bound states~\cite{wang2018single}. The dressed-photon scattering states therefore form a complete and orthonormal basis for the single-excitation subspace inside the light cone for each momentum $\bm{p}$. Details of the dressed-photon states, including their explicit form, as well as their orthonormality and completeness properties, are provided in Appendix~\ref{app:single_photon_states}.

It is important to emphasize that $t_{\bm{p}}(E)$ describes the scattering phase shift for the spatially symmetric photon mode, namely the equal superposition of photons propagating in the $+z$ and $-z$ directions, as previously described for 2D arrays. When a photon is incident from only one side of the array, its wavefunction can be decomposed into the symmetric mode $C^\dagger_{\bm{p}}(\chi)$, which couples to the atoms, and an orthogonal, spatially antisymmetric mode that does not couple. At the resonance of the collective mode, $\omega_{eg}+\Delta(\bm{p})$, the symmetric mode acquires a transmission coefficient $t_{\bm{p}}(E) = -1$, while the antisymmetric mode transmits with a coefficient of $+1$. The resulting interference leads to perfect reflection of the incident photon, which is the hallmark of atomic mirror behavior \cite{shahmoon2017cooperative,bettles2016enhanced, rui2020subradiant, srakaew2023subwavelength}.

\subsection{Effective spin model}

Our analysis of nonlinear scattering is built upon a well-established effective spin model whose dynamics are governed by a non-Hermitian Hamiltonian $H_{\mathrm{eff}}$. This effective model arises from the Markovian nature of the light-matter interaction, which permits the elimination of the photonic degrees of freedom. The resulting Hamiltonian acts only on the atomic subspace but contains all information required to describe the full system dynamics.
It is given by~\cite{ExponentialAsenjoGarcia2017}
\begin{equation}
H_{\mathrm{eff}} = \int_{\mathrm{BZ}} d\bm{p} \,   \epsilon(\bm{p})\, b^\dagger_{\bm{p}} b_{\bm{p}} +U,
\label{eqMintro}
\end{equation}
where the complex spin-wave dispersion $\epsilon(\bm{p})$ is given by 
\begin{equation}
\epsilon(\bm{p}) =
\begin{cases}
\Delta(\bm{p}) & \bm{p} \notin \text{LC} \\
\Delta(\bm{p}) - \frac{i}{2} \Gamma(\bm{p}) & \bm{p} \in \text{LC}
\end{cases}\label{eqepsilon}
\end{equation}
The dispersion is real for dark spin waves ($\bm{p} \notin \text{LC}$) and acquires an imaginary part for dissipative bright spin waves ($\bm{p} \in \text{LC}$). It can be obtained from a discrete Fourier transform of the free-space dipole-dipole interaction between emitters.

Crucially, all relevant photonic observables, including the S-matrix and scattering cross section, are determined directly from the atomic dynamics governed by this Hamiltonian. It therefore provides a complete and computationally efficient framework for the multi-excitation scattering problems considered in this work.

\section{Multi-excitation scattering matrix}
\label{sec_scatt_matrix}

In this section, we turn to the central topic of this work: multi-excitation scattering in the $n$-excitation subspace of the atomic array system. As established previously, the single-excitation eigenstates in the thermodynamic limit consist of dark spin waves and photons dressed by bright spin waves. However, when multiple excitations are present, the nonlinear interactions between atomic excitations give rise to much richer scattering physics.
The nonlinear interaction term in Eq.~\eqref{eq:U_interaction_intro} describes the process where two incoming atomic spin waves scatter into two outgoing atomic spin waves, with total lattice momentum $\bm{P}$ conserved but the relative momentum changing from $\bm{q}$ to $\bm{q}'$. Since photons are coupled to bright spin waves, this atomic interaction mediated by $U$ also gives rise to effective interactions among photons as well as between photons and atomic excitations.

We consider a \emph{finite} number of excitations, in which case scattering theory provides the natural framework for analysis. The central object is the $n$-excitation S-matrix, $S^{(n)}$, whose elements describe the transition amplitudes between asymptotically free incoming and outgoing states.

In our atomic array system, the elementary asymptotic excitations are photons and dark spin waves. Bright spin waves are transient states that decay into photons and thus do not form stable asymptotic states. We also note that for certain interaction potentials, multi-excitation dark-spin-wave bound states could form and act as additional stable scattering participants. In this work, however, we focus on the scattering between the elementary, unbound excitations. We therefore define our asymptotic states as consisting solely of individual photons and dark spin waves.
In experiments, incoming photons can be prepared using laser beams, while dark spin waves may be created through techniques such as rapid modulation of the lattice spacing, dynamic light shifts~\cite{Rubies-Bigorda2022}, or multi-photon excitation schemes~\cite{Rusconi2021}.

The asymptotic incoming and outgoing states, which contain $\alpha$ and $\beta$ photons, respectively, are thus defined as
\begin{subequations}
\begin{align}
|k^{(n,\alpha)}_{\mathrm{in}}\rangle &= \prod_{j=1}^{\alpha} C^\dagger_{\boldsymbol{p}_{j}}(\chi_j)\, \prod_{i=\alpha+1}^{n} b^\dagger_{\boldsymbol{p}_{i}} |0\rangle ,\label{eqPhiin}\\ 
|k^{(n,\beta)}_{\mathrm{out}}\rangle &= \prod_{j=1}^{\beta} C^\dagger_{\boldsymbol{p}'_{j}}(\chi'_j)\, \prod_{i=\beta+1}^{n} b^\dagger_{\boldsymbol{p}'_{i}} |0\rangle.
\label{eqPhiout}
\end{align}
\label{eqphiinout}
\end{subequations}
The lattice momenta distinguish the types of excitation according to the Light Cone (LC) criterion established earlier: momenta with indices $j$, $\bm{p}_j, \bm{p}'_j \in \text{LC}$ correspond to photons, while momenta with indices $i$, $\bm{p}_i, \bm{p}'_i \notin \text{LC}$ correspond to dark spin waves.

The S-matrix operator $S^{(n)}$ can be decomposed into blocks $S^{(n)}_{\alpha, \beta}$, which map the subspace of incoming states in channel $\alpha$ to the subspace of outgoing states in channel $\beta$. The matrix elements $\langle k^{(n,\beta)}_{\mathrm{out}} | S^{(n)}|k^{(n,\alpha)}_{\mathrm{in}}\rangle \equiv \langle k^{(n,\beta)}_{\mathrm{out}} | S_{\alpha,\beta}^{(n)}|k^{(n,\alpha)}_{\mathrm{in}}\rangle$ yield the probability amplitude for the corresponding scattering process. A formal definition of the S-matrix is provided in Section~\ref{sec_formal_S_matrix}.

\subsection{Relating the full S-matrix to the atomic T-matrix}
\label{sec_S_matrix_central}

In this section, we present our central result on the S-matrix, with the proof provided in Section~\ref{SubsecSmatrixProof}. The central result is a formula connecting the S-matrix $S^{(n)}$ to the T-matrix $\bar{T}^{(n)}(\omega)$ defined purely within the effective $n$-excitation atomic subspace. This relationship allows us to compute scattering observables in the full space by solving for the dynamics within the reduced atomic subspace.

First, we define the key components within the atomic subspace. Let $H_{\mathrm{eff}}^{(n)}$ and $U^{(n)}$ denote the restriction of $H_{\mathrm{eff}}$ [Eq.~\eqref{eqMintro}] and $U$ [Eq.~\eqref{eq:U_interaction_intro}] to the $n$-excitation subspace, respectively. Using these, we formally define the $n$-excitation atomic T-matrix $\bar{T}^{(n)}(\omega)$ via the Lippmann-Schwinger equation within this subspace:
\begin{equation}
\bar{T}^{(n)}(\omega)=U^{(n)}+U^{(n)}\frac{1}{\omega-H_{\mathrm{eff}}^{(n)}}\bar{T}^{(n)}(\omega). \label{eqTLS}
\end{equation}
This T-matrix encapsulates the full interaction dynamics mediated by the nonlinearity $U^{(n)}$ within the atomic subspace. The computation of $\bar{T}^{(n)}(\omega)$ follows a standard procedure in scattering theory, which we discuss in more detail in Section~\ref{subsectionTn}.

To express the connection between the S-matrix and $\bar{T}^{(n)}(\omega)$,  {we introduce auxiliary states defined entirely within the atomic subspace. These states are constructed by replacing all photons in the full incoming and outgoing states $|k^{(n,\alpha)}_{\mathrm{in}}\rangle$ and $|k^{(n,\beta)}_{\mathrm{out}}\rangle$ with bright spin waves, while the dark-spin-wave components remain unchanged. Specifically, these auxiliary states are labeled by the same sets of lattice momenta $\{\bm{p}_i\}$ and $\{\bm{p}'_i\}$ that appear in the asymptotic states of Eq.~\eqref{eqphiinout},
\begin{equation}
|p^{(n)}_{\mathrm{in}}\rangle  =\prod_{i=1}^{n}b^\dagger_{\boldsymbol{p}_{i}} \ket{0} \quad \text{and} \quad
|p^{(n)}_{\mathrm{out}}\rangle =\prod_{i=1}^{n}b^\dagger_{\boldsymbol{p}'_{i}} \ket{0}. \label{eqaa0} 
\end{equation}
We use the shorthand $p^{(n)}_{\mathrm{in}}=(\bm{p}_1, \dots, \bm{p}_n)$ and $p^{(n)}_{\mathrm{out}}=(\bm{p}'_1, \dots, \bm{p}'_n)$ to refer to these atomic spin wave momentum configurations.

Our main result, derived in Section~\ref{SubsecSmatrixProof}, relates the S-matrix element between the full asymptotic states to the atomic T-matrix:
\eq{
& \langle k^{(n,\beta)}_{\mathrm{out}} | S_{\alpha,\beta}^{(n)}|k^{(n,\alpha)}_{\mathrm{in}}\rangle =\langle k^{(n,\beta)}_{\mathrm{out}}|k^{(n,\alpha)}_{\mathrm{in}}\rangle    t(k^{(n,\alpha)}_{\mathrm{in}})\nonumber \\
 &  -2\pi i \delta(E-E')
\langle  p^{(n)}_{\mathrm{out}}|\bar{T}^{(n)}(E+i0) |p_{\mathrm{in}}^{(n)}\rangle   
  a(k^{(n,\alpha)}_{\mathrm{in}})  a(k^{(n,\beta)}_{\mathrm{out}}).
\label{eqSgen}
}Here, $t(k^{(n,\alpha)})$ denotes the collective transmission coefficient for the outgoing photons. It is defined as the product of the single-photon transmission coefficients [Eqs.~\eqref{eqepchi}] for each outgoing photon
\begin{equation}
t(k^{(n,\alpha)}) = \prod_{j=1}^{\alpha} t_{\bm{p}_j}(E_j), \label{eqtcollective}
\end{equation}
where $E_j=E_{\bm{p}_i}(\chi_i)$ is the energy of the photon with momentum \(\bm{p}_, \chi_j\).
Similarly, $a(k^{(n,\alpha)})$ is the collective atomic amplitude for the bright spin waves associated with the outgoing photons. It is given by the product over the individual atomic amplitudes [see Eq.~\eqref{eqt_pchi}]
\begin{equation}
a(k^{(n,\alpha)}) = \prod_{j=1}^{\alpha} a_{\bm{p}_j}
(E_j). \label{eqaknalpha}
\end{equation}

The first term in Eq.~(\ref{eqSgen}) represents the linear scattering part. It is diagonal in the momentum basis and describes processes where the excitations pass through the system without interacting with each other via the nonlinearity $U^{(n)}$. In this process, incoming photons simply acquire their total transmission coefficient $t(k^{(n,\alpha)}_{\mathrm{in}})$, while dark spin waves propagate without change.

The second term in Eq.~(\ref{eqSgen}) describes the nonlinear interaction dynamics, which are constrained by energy conservation enforced by the delta function $\delta(E-E')$, where $E$ and $E'$ denote the total energies of the incoming and outgoing states, respectively. This process unfolds in three steps, illustrated schematically in Fig.~\ref{fig:FeynmanDiag}(a): (i) the incoming photons in the state $|k^{(n,\alpha)}_{\mathrm{in}}\rangle$ are converted into atomic excitations with collective coupling amplitude $a(k^{(n,\alpha)}_{\mathrm{in}})$;  (ii) these excitations, together with any initial dark spin waves, interact within the atomic subspace, with the full dynamics described by the on-shell atomic T-matrix element $\langle p^{(n)}_{\mathrm{out}}|\bar{T}^{(n)}(E+i0)|p^{(n)}_{\mathrm{in}}\rangle$ \footnote{The term ``on-shell'' specifies that the T-matrix is evaluated at the complex frequency $\omega = E+i0$.
Here, $E$ corresponds to the total real energy of the system, and the infinitesimal imaginary part $+i0$ ensures the correct causal (outgoing-wave) boundary conditions are satisfied.}; and (iii) the resulting bright atomic excitations are converted back into outgoing photons in the state $|k^{(n,\beta)}_{\mathrm{out}}\rangle$, with an associated amplitude $a^*(k^{(n,\beta)}_{\mathrm{out}})$ [the complex conjugate of $a(k^{(n,\beta)}_{\mathrm{out}})$] and a linear transmission coefficient $t(k^{(n,\beta)}_{\mathrm{out}})$. 
They fulfill the relation $a^*(k^{(n,\beta)}_{\mathrm{out}}) t(k^{(n,\beta)}_{\mathrm{out}}) = a(k^{(n,\beta)}_{\mathrm{out}})$, simplifying the final form of Eq.~\eqref{eqSgen}.
In more general models, for example those involving additional decay channels where $t$ is not unitary, the term $a(k^{(n,\beta)}_{\mathrm{out}})$ in Eq.~\eqref{eqSgen} should be replaced by its original form $a^*(k^{(n,\beta)}_{\mathrm{out}})\, t(k^{(n,\beta)}_{\mathrm{out}})$.

We can examine some limiting cases of Eq.~\eqref{eqSgen}. If the nonlinear interaction is turned off ($U = 0$), then $U^{(n)}=0$ and consequently the atomic T-matrix $\bar{T}^{(n)}(\omega) = 0$. Only the first term in Eq.~\eqref{eqSgen} survives. This describes purely linear scattering where each incoming photon $(\bm{p}_j, \chi_j)$ acquires its transmission phase $t_{\boldsymbol{p}_j}(E_{\bm{p}}(\chi_j))$, and all dark spin waves pass through unaffected.

If the asymptotic states consist solely of dark spin waves ($\alpha = \beta = 0$), the coupling and transmission factors are absent. Then, Eq.~\eqref{eqSgen} simplifies to
\begin{equation}
\begin{split}
&\langle p^{(n)}_{\mathrm{out}}| S^{(n)} |p^{(n)}_{\mathrm{in}} \rangle = \langle p^{(n)}_{\mathrm{out}} | p^{(n)}_{\mathrm{in}} \rangle \\
&- 2\pi i \delta(E - E') \langle p^{(n)}_{\mathrm{out}} | \bar{T}^{(n)}(E + i0 ) | p^{(n)}_{\mathrm{in}} \rangle.
\end{split}
\label{eqSdarksw}
\end{equation}
This expression mirrors the standard relationship between the S-matrix and the T-matrix in potential scattering theory. However, its applicability here is non-trivial, since the conventional derivation assumes a Hermitian Hamiltonian. In our case, the effective atomic Hamiltonian $H_{\mathrm{eff}}^{(n)}$ is non-Hermitian because it incorporates the dissipative (\ie non-unitary) dynamics of bright spin waves, which are not true asymptotic states. The validity of Eq.~\eqref{eqSdarksw} in the dark-spin-wave sector demonstrates that the T-matrix, though derived from a non-Hermitian effective Hamiltonian, still correctly describes the physical scattering processes between the true, non-dissipative asymptotic states.

\begin{figure}[t]
    \centering
\includegraphics[width=1\linewidth]{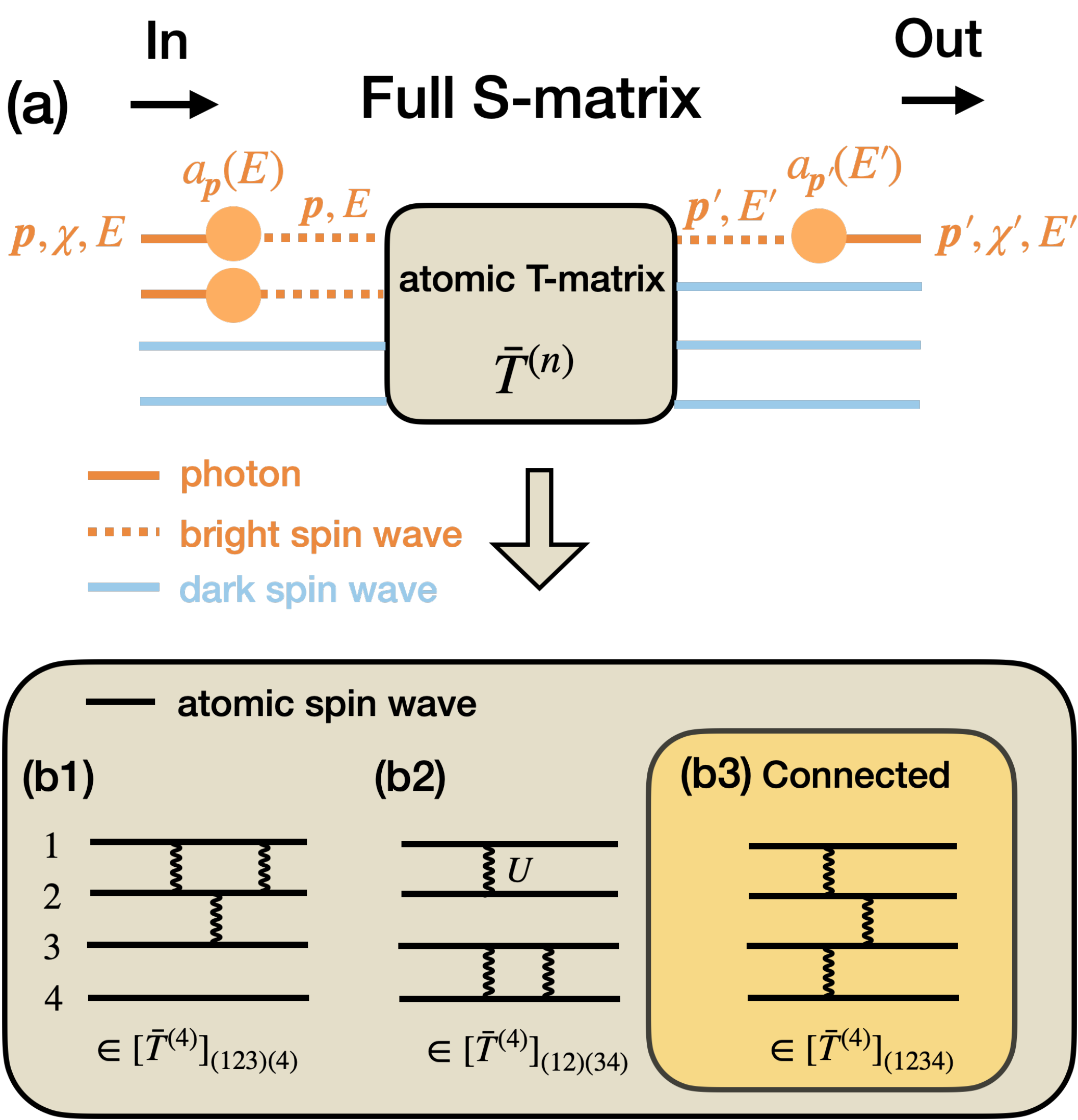}
    \caption{\textbf{Mapping the full S-matrix to the atomic T-matrix.} 
    \textbf{(a)} This panel diagrammatically represents our main result [Eq.~\eqref{eqSgen}], showing how the interacting part of the S-matrix is determined by the atomic T-matrix \(\bar{T}^{(n)}\). The calculation proceeds in three steps: (i) Each incoming photon \((\bm{p}, \chi)\) is mapped to a bright spin wave with an amplitude \(a_{\bm{p}}(E)\), where \(E=E_{\bm{p}}(\chi)\) is the photon's energy. (ii) These excitations, along with any initial dark spin waves, interact within the atomic subspace via \(\bar{T}^{(n)}\). (iii) Every resulting bright spin wave \(\bm{p}'\) is mapped back to an outgoing photon \((\bm{p}', \chi')\) with an amplitude \(a_{\bm{p}'}(E')\), where $E'=E_{\bm{p}'}(\chi')$ is the energy of the output photon. Within our model, this outgoing amplitude is related to the single-photon transmission coefficient \(t_{\bm{p}'}\) via \(a_{\bm{p}'}(E') = a^*_{\bm{p}'}(E') t_{\bm{p}'}(E')\).
    \textbf{(b1--b3)} Example diagrams illustrating the cluster decomposition of the atomic T-matrix. Panel~(b3) shows a fully connected diagram where all excitations interact simultaneously.}
    \label{fig:FeynmanDiag}
\end{figure}

\subsection{Computing the atomic T-matrix $\bar{T}^{(n)}$\label{subsectionTn}}

There is a standard procedure for computing the atomic $n$-excitation T-matrix, $\bar{T}^{(n)}$. The key idea is that its connected component can be constructed iteratively from lower-excitation T-matrices. This relies on a decomposition of $\bar{T}^{(n)}(\omega)$ into a sum over different clusters of lower-excitation T-matrices.

This decomposition is known as a \textit{cluster decomposition}~\cite{Glockle2012}, and takes the general form
\begin{equation}
\bar{T}^{(n)}(\omega) = \sum_{\zeta \in \mathcal{P}_n} [\bar{T}^{(n)}]_{\zeta}, 
\label{eqTnDecomp}
\end{equation}
where $\mathcal{P}_n$ denotes the set of all partitions of the $n$ excitations into disjoint, non-empty subsets, called clusters. Each term $[\bar{T}^{(n)}]_{\zeta}$ corresponds to a situation in which excitations within each cluster interact among themselves, without interacting with excitations in other clusters.

For example, with $n=3$ excitations, there are five possible partitions: $(1)(2)(3)$ (all excitations isolated), $(12)(3)$, $(13)(2)$, $(23)(1)$ (one pair interacts), and $(123)$ (all excitations interact). Each corresponding term in the sum contains diagrams where the specified clusters are internally connected through interactions, but disconnected from each other. Figure~\ref{fig:FeynmanDiag}(b1-b3) depicts examples of Feynman diagrams included in three different clusters of the four-excitation T-matrix.

The most important piece is the \emph{connected} part of the T-matrix, denoted $\bar{T}^{(n)}_c(\omega)$, which corresponds to the contribution from the partition where all $n$ excitations belong to a single cluster
\begin{equation}
\bar{T}_c^{(n)}(\omega) = [\bar{T}^{(n)}]_{(12\dots n)}.
\end{equation}
This contains all diagrams in which all $n$ excitations are connected by interaction lines, and thus represents a genuine $n$-excitation scattering process [see the $n=4$ example in Fig.~\ref{fig:FeynmanDiag}(b3)]. All other partitions in Eq.~\eqref{eqTnDecomp} can be viewed as tensor products of lower-order connected T-matrices. For instance,
\begin{equation}
[\bar{T}^{(5)}]_{(13)(245)} = [\bar{T}^{(2)}]_{(13)} \otimes [\bar{T}^{(3)}_c]_{(245)},
\end{equation}
describes a process where excitations 1 and 3 interact via a two-body T-matrix, while excitations 2, 4, and 5 participate in a separate connected three-body process.

Given the connected T-matrices for $j < n$, the $n$-excitation connected T-matrix $\bar{T}^{(n)}_c$ can be computed from the Rosenberg equations \cite{Rosenberg1965}, which form a system of $n(n-1)$ coupled integral equations. After computing $\bar{T}_c^{(n)}$, the full T-matrix $\bar{T}^{(n)}$ is obtained via the cluster decomposition in Eq.~\eqref{eqTnDecomp}.

Since the interaction $U^{(n)}$ conserves total lattice momentum, the atomic T-matrix must also respect momentum conservation. Thus, its matrix elements take the form
\begin{equation}
\begin{aligned}
&\langle p^{(n)}_{\mathrm{out}} | \bar{T}_c^{(n)}(E+i0) | p^{(n)}_{\mathrm{in}} \rangle \\
&= \delta(\bm{P} - \bm{P}') \langle p^{(n)}_{\mathrm{out}} | \bar{T}_c^{(n)}(E+i0, \bm{P}) | p^{(n)}_{\mathrm{in}} \rangle,
\end{aligned}
\label{eqTbarfixP}
\end{equation}
where $\bm{P}$ and $\bm{P}'$ are the total lattice momenta of the incoming and outgoing states, respectively. We refer to $\bar{T}_c^{(n)}(E+i0, \bm{P})$ as the connected T-matrix evaluated at fixed total momentum $\bm{P}$.

 \subsection{Derivation of the S-matrix}
\label{SubsecSmatrixProof}
In this section, we derive our central result on the S-matrix, i.e. Eq.~\eqref{eqSgen}, starting from the formal definition of the S-matrix.
\subsubsection{Formal definition of the S-matrix}
\label{sec_formal_S_matrix}

To define the S-matrix, it is essential to distinguish between the ``bare" part of the Hamiltonian, which accounts for the free propagation of excitations, and the ``interacting" part, which facilitates scattering. We decompose the Hamiltonian \(H\) in Eq.~\eqref{eqHamilSimp} into three components:
\begin{subequations}
\begin{align}
H &= H_0 + V + U, \\
H_0 &= \int_{\mathrm{LC}} d\bm{p}\ H_{\mathrm{photon}, \bm{p}} + \int_{\bm{p} \notin \mathrm{LC}} d\bm{p}\ H_{\mathrm{atom}, \bm{p}}, \\
V &= \int_{\mathrm{LC}} d\bm{p}\ V_{\bm{p}}+ \int_{\bm{p} \in \mathrm{LC}} d\bm{p}\ H_{\mathrm{atom}, \bm{p}}, 
\end{align}
\end{subequations}
where 
\begin{subequations}
\begin{align}
H_{\text{photon}, \boldsymbol{p}} &= \int_0^{+\infty} d\chi \, E_{\boldsymbol{p}}(\chi) C^\dagger_{\boldsymbol{p}}(\chi) C_{\boldsymbol{p}}(\chi), \\
H_{\text{atom}, \boldsymbol{p}} &= \Delta(\boldsymbol{p}) b^\dagger_{\boldsymbol{p}} b_{\boldsymbol{p}}, \\
V_{\boldsymbol{p}} &= \int_0^{+\infty} d \chi \, [g_{\boldsymbol{p}} C^\dagger_{\boldsymbol{p}}(\chi) b_{\boldsymbol{p}} + \text{h.c.}],
\end{align}
\end{subequations}
Here, \(H_0\) represents the energy of free photons and dark spin waves. Since we are interested in the scattering processes of these excitations, \(H_0\) serves as the ``bare" part of the Hamiltonian, characterizing the propagation of ``free excitations" both before and after interactions occur. 
The terms \(V + U\) comprise the ``interacting" part of the Hamiltonian: \(V\) includes the energy of the bright spin waves and the quadratic coupling between photons and bright spin waves, while \(U\) accounts for the nonlinear interactions among atoms.

In the context of the \(n\)-excitation subspace, we denote operators pertinent to this subspace with the superscript \((n)\). The S-matrix is defined in terms of the Møller operators \(\Omega^{(n)}_\pm\)
\begin{subequations}
\begin{align}
S^{(n)} &\equiv \Omega^{(n)\dagger}_- \Omega^{(n)}_+, \\
\Omega^{(n)}_\pm &\equiv \lim_{t \rightarrow \mp \infty} \exp(iH^{(n)}t)\exp(-iH^{(n)}_0 t).
\end{align}
\label{eqSdefi1}
\end{subequations}
The Møller operators are constructed from the infinite time evolution under the bare Hamiltonian, counterposed with the time evolution under the full Hamiltonian. Specifically, \(\Omega^{(n)}_\pm\) relate the eigenstates of \(H^{(n)}\) to their input and output asymptotic states, which are aligned with the eigenstates of \(H^{(n)}_0\).

In standard scattering theory, the S-matrix is related to the T-matrix \( T^{(n)}(E+i0) \), which can be calculated using the Lippmann-Schwinger equations
\begin{subequations}
\begin{align}
S^{(n)} &= \mathbb{1}^{(n)} - 2\pi i \delta(E-E') T^{(n)}(E+i0), \label{eqSformalT}\\
T^{(n)}(\omega) &= [U^{(n)} + V^{(n)}] + [U^{(n)} + V^{(n)}]\frac{1}{\omega - H_0^{(n)}} T^{(n)}(\omega). \label{eqTUVLS}
\end{align}
\end{subequations}
Here, \(\mathbb{1}^{(n)}\) denotes the identity matrix in the \( n \)-excitation subspace, while \( E \) and \( E' \) are the energies of the incoming and outgoing states, respectively.

However, solving Eq.~\eqref{eqTUVLS} directly is challenging due to the interaction term \( U^{(n)} + V^{(n)} \), which involves both quadratic and quartic interactions and includes operators for both atoms and photons.

To address this complexity, we adopt an alternative approach. By re-partitioning the full Hamiltonian into different ``bare" and ``interacting" components, we can define a new S-matrix, \( S_U^{(n)} \), which is easier to solve. After determining \( S_U^{(n)} \), we can establish its relationship to \( S^{(n)} \) to obtain the desired results.

\subsubsection{S-matrix for dressed photons}
\label{AppSecSmatrixdressed}

To define the S-matrix \(S_U^{(n)}\), we treat \(H_0 + V\) as the bare Hamiltonian and consider \(U\) as the interaction part responsible for scattering between the eigenstates of \(H_0 + V\). This is expressed as follows:
\begin{align}
S_U^{(n)} &= \Omega_{U-}^{(n)\dagger} \Omega^{(n)}_{U+}, \label{eqSdressed} \\
\Omega^{(n)}_{U\pm} &= \lim_{t \rightarrow \mp \infty} \exp(iH^{(n)}t) \exp[-i(H^{(n)}_{0} + V^{(n)})t].
\end{align}
The eigenstates of \(H_0 + V=\int_{\mathrm{BZ}} d\bm{p}\ H_{\bm{p}}\) are the dark spin waves and the dressed photons, as described in Section~\ref{sub_sec_single_excitation}. Thus, this S-matrix $S_U^{(n)}$ characterizes scattering not  between \emph{free photons} and dark spin waves, but rather between \emph{dressed photons} and dark spin waves. We refer to $S_U^{(n)}$ as the \emph{dressed-excitation S-matrix}, to distinguish it from the \emph{full S-matrix}, $S^{(n)}$, which is the primary quantity computed in this work.

It is apparent from the definitions of the two S-matrices that they are related by the equation
\begin{subequations}
\begin{align}
S^{(n)} &= \Omega_{0-}^{(n)\dagger} S_U^{(n)} \Omega^{(n)}_{0+}, \label{eqAppendixSSu} \\
\Omega_{0\pm} &\equiv \lim_{t \rightarrow \mp \infty} \exp[i(H_0 + V)t] \exp[-iH_{0}t].
\end{align}
\end{subequations}
Here, \(\Omega_{0\pm}\) are the Møller operators when treating \(H_0 + V\) as the full Hamiltonian and \(H_0\) as the bare Hamiltonian and $\Omega_{0\pm}^{(n)}$ are their restrictions to the $n$-excitation subspace. 
These operators connect free photons to dressed photons according to
\begin{subequations}
\begin{align}
\Omega_{0+} C^\dagger_{\bm{p}}(\chi) |0\rangle &= \psi^\dagger_{\bm{p}}(\chi)|0\rangle, \\
\langle 0| C_{\bm{p}}(\chi) \Omega_{0-}^\dagger &= \langle 0| \psi_{\bm{p}}(\chi) t_{\bm{p}}(\chi),
\end{align}
\label{eqSPMollarRelations}
\end{subequations}
and do not act on the dark spin waves.
Accordingly, we define the linear-scattering S-matrix, which describes single-excitation scattering, as
\begin{equation}
S^{(n)}_{\mathrm{linear}}\equiv \Omega^{(n)\dagger}_{0-} \Omega^{(n)}_{0+}. \label{eqSlinear} 
\end{equation}
The action of this operator on an incoming state [Eq.~\eqref{eqPhiin}] is to impart the single-excitation transmission coefficient to each photon component while acting as the identity on all dark-spin-wave components. In the single-excitation subspace this reduces to the full S-matrix: $S^{(1)} = S^{(1)}_{\mathrm{linear}}$.

Given the input and output states defined in Eq.~\eqref{eqphiinout}, we introduce corresponding states where each free photon is replaced by a dressed photon with the same momentum
\begin{subequations}
\begin{align}
|\psi^{(n,\alpha)}_{\mathrm{in}}\rangle &= \prod_{i=1}^{\alpha} \psi^\dagger_{\boldsymbol{p}_i}(\chi_i) \prod_{j=\alpha+1}^{n} b^\dagger_{\boldsymbol{p}_j} |0\rangle , \\
|\psi^{(n,\beta)}_{\mathrm{out}}\rangle &= \prod_{i=1}^{\beta}  \psi^\dagger_{\boldsymbol{p}'_i}(\chi'_i)\prod_{j=\beta+1}^{n} b^\dagger_{\boldsymbol{p}'_j}|0\rangle.
\end{align}
\label{eqAppendixdressedInOut}
\end{subequations}
From Eqs.~\eqref{eqSPMollarRelations} and \eqref{eqAppendixSSu}, we have
\begin{equation}
\langle k^{(n,\beta)}_{\mathrm{out}}| S^{(n)}|k^{(n,\alpha)}_{\mathrm{in}}\rangle = \langle \psi^{(n,\beta)}_{\mathrm{out}}| S_{U}^{(n)}|\psi^{(n,\alpha)}_{\mathrm{in}}\rangle \, t(k^{(n,\beta)}_{\mathrm{out}}),\label{eqAppendixSnSu}
\end{equation}
where the collective transmission coefficient $t(k^{(n,\beta)}_{\mathrm{out}})$ is defined in Eq.\ \eqref{eqtcollective}.
The remaining task is to derive \(\langle \psi^{(n,\beta)}_{\mathrm{out}}| S_{U}^{(n)}|\psi^{(n,\alpha)}_{\mathrm{in}}\rangle\).
Following the standard formalism of scattering theory, the dressed-excitation S-matrix $S_{U}^{(n)}$ is related to the dressed-excitation T-matrix \(T_U^{(n)}(E+i0)\), as defined by the Lippmann-Schwinger equation
\begin{subequations}
\begin{align}
&\langle \psi^{(n,\beta)}_{\mathrm{out}}| S_{U}^{(n)}|\psi^{(n,\alpha)}_{\mathrm{in}}\rangle = \langle  
\psi^{(n,\beta)}_{\mathrm{out}}|\psi^{(n,\alpha)}_{\mathrm{in}}\rangle \nonumber\\
&- 2\pi i \delta(E-E') \langle \psi^{(n,\beta)}_{\mathrm{out}}|T_U^{(n)}(E+i0)|\psi^{(n,\alpha)}_{\mathrm{in}}\rangle, \label{eqSmT}
\end{align}
\begin{equation}
T_U^{(n)}(\omega) = U^{(n)} + U^{(n)}\frac{1}{\omega-H_0^{(n)}-V^{(n)}}T_U^{(n)}(\omega). \label{AppenTUnLS}
\end{equation}
\end{subequations}
Here, \(E\) and \(E'\) denote the total energies of the incoming and outgoing states.

Since \(U^{(n)}\) is an operator acting on atomic excitations, the T-operator \(T^{(n)}_U\) generated by \(U^{(n)}\) also acts on the atomic subspace. Given our Markovian system, for any \(n\)-excitation atomic spin wave \(|p^{(n)}\rangle = \prod_{i=1}^{n} b^\dagger_{\boldsymbol{p}_i} |0\rangle\),
\begin{equation}
\braket{p^{(n)}|\frac{1}{\omega-H_0^{(n)}-V^{(n)}}|p^{(n)}} = \braket{p^{(n)}|\frac{1}{\omega-H_{\mathrm{eff}}^{(n)}}|p^{(n)}}. \label{eqAppbpropa}
\end{equation}

Using Eq.~\eqref{eqAppbpropa} and comparing Eq.~\eqref{AppenTUnLS} with Eq.~\eqref{eqTLS}, we deduce that
\begin{equation}
T_U^{(n)}(\omega) = \bar{T}^{(n)}(\omega) \otimes \mathbb{1}^{(n)}_{\mathrm{ph}}, \label{eqTueqTbar}
\end{equation}
where \(\mathbb{1}^{(n)}_{\mathrm{ph}}\) is the identity operator on the photon subspace.

Using Eq.~\eqref{eqTueqTbar} and the identity
\(
\langle 0|b_{\boldsymbol{p}} |\psi_{\boldsymbol{p}}(\chi)\rangle = a_{\boldsymbol{p}}(\chi)
\) [see Eq.~\eqref{eqdressedPh_app}],
Eq.~\eqref{eqSmT} becomes
\eq{
&\langle \psi^{(n,\beta)}_{\mathrm{out}} | S_U^{(n)}|\psi^{(n,\alpha)}_{\mathrm{in}}\rangle = \langle \psi^{(n,\beta)}_{\mathrm{out}}|\psi^{(n,\alpha)}_{\mathrm{in}}\rangle \nonumber\\
&- 2\pi i \delta(E-E') \langle p^{(n)}_{\mathrm{out}}|\bar{T}^{(n)}(E+i0) |p_{\mathrm{in}}^{(n)}\rangle  
 a(k^{(n,\alpha)}_{\mathrm{in}}) a^*(k^{(n,\beta)}_{\mathrm{out}}).
\label{AppendixeqSgen}
}

Together with Eq.~\eqref{eqAppendixSnSu} and the identity \(a^*(k^{(n,\beta)}_{\mathrm{out}}) \cdot t(k^{(n,\beta)}_{\mathrm{out}})=a(k^{(n,\beta)}_{\mathrm{out}})\), this results in Eq.~\eqref{eqSgen}, our main result on the full S-matrix.

\section{Scattering cross section\label{sec_cs}}

In this section, we introduce the \( n \)-body scattering cross section $\sigma^{(n)}$, a central experimental observable that quantifies the efficiency of multi-excitation scattering processes. While the S-matrix encodes the probability amplitudes for transitions between specific momentum states, the cross section measures the number of excitations scattered per unit time and per unit incident flux. For a given input channel $\alpha$, the total cross section $\sigma_{\alpha,\mathrm{tot}}^{(n)}(k^{(n,\alpha)})$ enables direct comparison of scattering strengths across different incoming states $k^{(n,\alpha)}$, while the partial cross sections $\sigma_{\alpha,\beta}^{(n)}(k^{(n,\alpha)})$ resolve the branching ratios for scattering into distinct output channels $\beta$.
\subsection{Definition} \label{SecCrossSectionDefi}

The scattering cross section is derived from the interacting part of the S-matrix. As shown in Eq.~\eqref{eqSgen}, the S-matrix consists of a linear term and a nonlinear term proportional to the atomic T-matrix, $\bar{T}^{(n)}$. However, $\bar{T}^{(n)}$ accounts for all possible scattering events, including those in which only a subset of the $n$ excitations interact (see the cluster decomposition in Sec.~\ref{subsectionTn}). To define a cross section that specifically measures the rate of true $n$-excitation scattering, we must isolate the contribution proportional to the \emph{connected} part of the atomic T-matrix, $\bar{T}_c^{(n)}$. We thus define the connected T-matrix, $T_c^{(n)}$, whose matrix elements capture the amplitude for genuine $n$-excitation events involving both photons and dark spin waves:

\begin{equation}
\begin{split}
\langle k^{(n,\beta)}_{\mathrm{out}}| T_c^{(n)}(\omega, \bm{P})|k^{(n,\alpha)}_{\mathrm{in}}\rangle = 
\langle p^{(n)}_{\mathrm{out}}|\bar{T}_c^{(n)}(\omega,\bm{P}) |p_{\mathrm{in}}^{(n)}\rangle \\
\times a(k^{(n,\alpha)}_{\mathrm{in}})\, a(k^{(n,\beta)}_{\mathrm{out}}).
\end{split}
\label{eqSgenT}
\end{equation}

The partial cross section for scattering from a specific incoming state $|k^{(n,\alpha)}_{\mathrm{in}}\rangle$ into the final channel $\beta$ is defined as
\begin{equation}
\begin{split}
\sigma_{\alpha,\beta}^{(n)}(k^{(n,\alpha)}_{\mathrm{in}}) = 4\pi^2 v_g(k^{(n,\alpha)}_{\mathrm{in}})^{-1}\int_{\mathcal{M}_\beta^{(n)}} d k^{(n,\beta)}_{\mathrm{out}}
\, \delta(E - E(k^{(n,\beta)}_{\mathrm{out}})) \\
\times \delta(\bm{P}-\bm{P}(k^{(n,\beta)}_{\mathrm{out}})) 
|\langle k^{(n,\beta)}_{\mathrm{out}}|T_c^{(n)}(E+i0, \bm{P})|k^{(n,\alpha)}_{\mathrm{in}}\rangle|^2,
\end{split}
\label{eq_cs_inte_n}
\end{equation}
where $v_g(k^{(n,\alpha)}_{\mathrm{in}})$ is the group velocity of the incoming state.

We use the shorthand notation $k^{(n,\alpha)}_{\mathrm{in}}=(\bm{p}_1, \dots, \bm{p}_n, \chi_1, \dots, \chi_{\alpha})^T$ and $k^{(n, \beta)}_{\mathrm{out}}=(\bm{p}'_1, \dots, \bm{p}'_n, \chi'_1, \dots, \chi'_{\beta})^T$ for the set of momenta. Here, $E(k^{(n,\beta)}_{\mathrm{out}})$ and $\bm{P}(k^{(n,\beta)}_{\mathrm{out}})=\sum_{j=1}^n\bm{p}'_j$ are the total energy and lattice momentum of the outgoing state, respectively.

To compute the group velocity, let $l^{(n-1,\alpha)}$ denote the $n-1$ mutually orthogonal lattice momenta and transverse wavenumbers perpendicular to the total momentum $\bm{P}$. The group velocity is then given by
\begin{equation}
v_g(k^{(n,\alpha)}_{\mathrm{in}}) = \left\| \nabla_{l^{(n-1,\alpha)}} E(k^{(n,\alpha)}_{\mathrm{in}}) \right\|,
\end{equation}
where the gradient is evaluated at fixed total momentum $\bm{P}$.

The integration in Eq.~\eqref{eq_cs_inte_n} is performed over all allowed momentum configurations of the final state, comprising $\beta$ photons and $n-\beta$ dark spin waves. Explicitly, the integration measure is defined as
\begin{align}
&\int_{\mathcal{M}_{\beta}^{(n)}} d k^{(n,\beta)}_{\mathrm{out}}   \nonumber \\
&\equiv 
\int_{\mathcal{D}_{\mathrm{bright}}^{(\beta)}} \prod_{l=1}^{\beta} d\bm{p}'_l  
\int_{-\infty}^{+\infty} \prod_{j=1}^{\beta} d\chi'_j\;
\int_{\mathcal{D}_{\mathrm{dark}}^{(n-\beta)}} \prod_{j=\beta+1}^n d\bm{p}'_j,
\end{align}
where $\mathcal{D}_{\mathrm{bright}}^{(\beta)}$ and $\mathcal{D}_{\mathrm{dark}}^{(n-\beta)}$ are the domains of lattice momenta for the bright and dark spin waves, respectively. We impose an ordering on these momenta to ensure correct symmetrization for indistinguishable excitations.

In one dimension, the domains are given by
\begin{subequations}
\eq{
    \mathcal{D}_{\mathrm{bright}}^{(\beta)} &= \{ p_j < p_{j+1} \mid p_j \in \mathrm{LC},~1 \leq j \leq \beta\}, \\
    \mathcal{D}_{\mathrm{dark}}^{(n-\beta)} &= \{p_j < p_{j+1} \mid p_j \notin \mathrm{LC},~\beta+1 \leq j \leq n\}.
}
\end{subequations}
In two dimensions, ordering is imposed on the $x$-components:
\begin{subequations}
\begin{align}
    \mathcal{D}_{\mathrm{bright}}^{(\beta)} &= \{ p_{j,x} < p_{j+1,x} \mid \bm{p}_j \in \mathrm{LC},~1 \leq j \leq \beta \}, \\
    \mathcal{D}_{\mathrm{dark}}^{(n-\beta)} &= \{ p_{j,x} < p_{j+1,x} \mid \bm{p}_j \notin \mathrm{LC},~\beta+1 \leq j \leq n \}.  
\end{align}
\end{subequations}

The total cross section for an incoming state $|k^{(n,\alpha)}_{\mathrm{in}}\rangle$ is given by
\begin{equation}
\sigma_{\alpha,\mathrm{tot}}^{(n)}(k^{(n,\alpha)}_{\mathrm{in}}) = \sum_{\beta=0}^n \sigma_{\alpha,\beta}^{(n)}(k^{(n,\alpha)}_{\mathrm{in}}). 
\end{equation}

We identify the characteristic length scaling of the scattering cross sections $\sigma_{\alpha,\mathrm{tot}}^{(n)}$, given that single atoms have a decay rate $\Gamma_0$ and the atomic nonlinearity has an effective range $\xi$. For two-level nonlinearities, $\xi \sim d$, where $d$ is the lattice spacing; for Rydberg nonlinearities, $D$ corresponds to the blockade radius \cite{Saffman2010}.
For 2D atomic arrays, the scattering cross sections scale as
\begin{equation}
\sigma_{\alpha,\mathrm{tot}}^{(n)} \sim 
\begin{cases}
\xi^{2n-3} & \alpha = 0 \\
\xi^{2(n-1)} (c/\Gamma_0)^{\alpha-1} & \alpha = 1, \dots, n
\end{cases},
\end{equation}
and for a 1D array,
\begin{equation}
\sigma_{\alpha,\mathrm{tot}}^{(n)} \sim
\begin{cases}
\xi^{n-2} & \alpha = 0 \\
\xi^{n-1}(c/\Gamma_0)^{\alpha-1}  & \alpha = 1, \dots, n
\end{cases}.
\end{equation}

Focusing on the specific case of two-excitation scattering from a pair of dark spin waves (\(\alpha=0\)) in a 1D array, the scattering cross section is dimensionless. Since the scattering process is unitary, \(\sigma^{(2)}_{0,0}\) and \(\sigma^{(2)}_{0,\mathrm{tot}}\) are both upper-bounded by 4.\footnote{This upper bound is a general property of unitary scattering when the cross section is dimensionless. For a normalized incoming state $|\psi_{\mathrm{in}}\rangle$, the scattered part of the wavefunction is $|\psi_{\mathrm{sc}}\rangle = |\psi_{\mathrm{out}}\rangle - |\psi_{\mathrm{in}}\rangle$. The triangle inequality then gives the total scattering probability as $\sigma_{\text{tot}} = \||\psi_{\mathrm{sc}}\rangle\|^2 \le (\||\psi_{\mathrm{out}}\rangle\| + \||\psi_{\mathrm{in}}\rangle\|)^2 = (1+1)^2 = 4$.} The remaining partial cross sections, \(\sigma^{(2)}_{0,1}\) and \(\sigma^{(2)}_{0,2}\), represent the probability of scattering into final states containing one and two photons, respectively. We illustrate these quantities in a simple example in Section~\ref{sec:1d_example}.

\subsection{Relating the scattering cross section to the atomic T-matrix\label{SubsecCsmainresult}}

In this subsection, we present our main results for the $n$-excitation scattering cross sections; the detailed derivation is given in Sec.~\ref{secDerScatteringCS}. The key result is that all integrals over the orthogonal photon wavenumbers $\chi_j$ can be evaluated analytically, allowing the partial cross section from Eq.~\eqref{eq_cs_inte_n} to be expressed entirely in terms of integrals over lattice momenta. The resulting expression for the partial cross section is
\begin{widetext}
\begin{equation}
\begin{split}
\sigma^{(n)}_{\alpha,\beta} (k^{(n,\alpha)}_{\mathrm{in}}) = 4\pi v_g(k^{(n,\alpha)}_{\mathrm{in}})^{-1}\, a(k^{(n,\alpha)}_{\mathrm{in}})
\int_{\mathcal{D}_{\mathrm{bright}}^{(\beta)}} \prod_{j=1}^{\beta} d\bm{p}'_j
\int_{\mathcal{D}_{\mathrm{dark}}^{(n-\beta)}} \prod_{j=\beta+1}^n d\bm{p}'_j\;
\left[
-\mathrm{Im}\frac{1}{E + i0 - \epsilon^{(n)}(p^{(n)}_{\mathrm{out}})}
\right] \\
\times \delta\big(\bm{P} - \bm{P}(p^{(n)}_{\mathrm{out}})\big)
\left|
\langle p^{(n)}_{\mathrm{out}} | \bar{T}_c^{(n)}(E + i0, \bm{P}) | p^{(n)}_{\mathrm{in}} \rangle
\right|^2,
\end{split}
\label{eqCrossSectionMain}
\end{equation}
\end{widetext}
where $\epsilon^{(n)}(p^{(n)}_{\mathrm{out}}) = \sum_{j=1}^n \epsilon(\bm{p}'_j)$ is the total energy of the outgoing atomic spin waves.

This formulation reduces the computation of the partial cross section to an integration over only the lattice momentum variables, with all transverse wavenumber integrals evaluated analytically. As a result, Eq.~\eqref{eqCrossSectionMain} provides a compact and practical starting point for analyzing $n$-excitation scattering, given knowledge of the connected T-matrix elements.

\subsection{Derivation of the scattering cross section}
\label{secDerScatteringCS}

In this section, we derive the main result for the $n$-excitation scattering cross section, given in Eq.~\eqref{eqCrossSectionMain}, starting from the definition in Eq.~\eqref{eq_cs_inte_n}.

The derivation begins by noting that the squared matrix element of the connected T-matrix for free photons can be written as
\begin{equation}
\begin{split}
\left| \langle k^{(n,\beta)}_{\mathrm{out}} | T_c^{(n)}(\omega, \bm{P}) | k^{(n,\alpha)}_{\mathrm{in}} \rangle \right|^2 = \left| \langle p^{(n)}_{\mathrm{out}} | \bar{T}_c^{(n)} (\omega, \bm{P}) | p^{(n)}_{\mathrm{in}} \rangle \right|^2 \\
\times \prod_{j=1}^{\alpha} |a_{\bm{p}_j}(E_j)|^2 \prod_{j=1}^{\beta} |a_{\bm{p}'_j}(E'_j)|^2,
\end{split}
\end{equation}
where we have used Eq.~\eqref{eqSgenT}, and $a_{\bm{p}_j}(E_j)$ and $a_{\bm{p}'_j}(E'_j)$ denote the atomic amplitudes [Eq.~\eqref{eqepchi}] for  photons with energies $E_j=E_{\bm{p}_j}(\chi_{j})$ and $E'_j=E_{\bm{p}'_j}(\chi'_{j})$, respectively.  

A key step is the analytical evaluation of the integrals over the $\beta$ orthogonal photon wavenumbers $\chi'_j$ appearing in Eq.~\eqref{eq_cs_inte_n}.  In Appendix \ref{appenKeyIdentity}, we derive that, under the Markov approximation,
\begin{align}
&\pi\int d\chi'_1 \cdots d\chi'_\beta~ \delta\left(E - \sum_{j=1}^\beta  E'_j \right) \prod_{j=1}^\beta |a_{\bm{p}'_j}(E'_j)|^2 \nonumber \\
&= -\mathrm{Im} \frac{1}{E + i0 - \sum_{j=1}^\beta \epsilon(\bm{p}'_j)}, \label{eqintegral_identity}
\end{align}
 where $E'_j=E_{\bm{p}'_j}(\chi'_{j})$. 
 
Applying this relation to Eq.~\eqref{eq_cs_inte_n} yields the compact expression for the partial cross section shown in Eq.~\eqref{eqCrossSectionMain}, thus completing the derivation.

\subsection{Relation to coherent beam transmission}

In this subsection, we demonstrate how the transmission and reflection of a coherent-state photon beam passing through a 2D atomic array are affected by nonlinear interactions. We present a perturbative relation between the transmission and reflection of the beam and the $m$-photon total scattering cross sections $\sigma^{(m)}_{m, \mathrm{tot}}$.

Consider the scenario where a weak Gaussian beam traverses the 2D atomic array. We assume that the cross section of the beam is significantly larger than the lattice spacing, so any photon within the beam can be approximated as a plane wave with a well-defined momentum. For a single photon, the transmission coefficient $t$ and the reflection coefficient $r$ can be calculated using the methods outlined in Section \ref{sub_sec_single_excitation}.

We aim to explore the impact of nonlinear atomic interactions on the beam's transmission and reflection. Let $R$ represent the incident photon flux (photons per unit time per unit area). The total rate of $m$-photon scattering events per unit time and area is given by $ \sigma^{(m)}_{m, \mathrm{tot}} R^m / m!$, where the $1/m!$ accounts for the indistinguishability of the photons in the incident beam. Since each such event removes $m$ photons, the total rate of photon loss per unit time and area due to $m$-photon processes is $m \times \frac{1}{m!} \sigma^{(m)}_{m, \mathrm{tot}} R^m = \frac{1}{(m-1)!} \sigma^{(m)}_{m, \mathrm{tot}} R^m$. The probability that a single ``test" photon is lost to such a process corresponds to this rate divided by the incident flux $R$.

Summing over all possible multi-photon scattering events, the probability $P_r$ that a given photon in the coherent state input does \emph{not} undergo a nonlinear scattering event is given by (see Appendix \ref{app:coherent_transmission} for a full derivation)
\begin{equation}
P_r = 1 - \sum_{m=2} \frac{1}{(m-1)!} \sigma^{(m)}_{m, \mathrm{tot}} R^{m-1}. \label{eqPsurvival}
\end{equation}
Assuming that photons experiencing nonlinear scattering are deflected out of the detection direction, the total transmitted and reflected photon fluxes are given by the single-photon results, reduced by this survival probability $P_r$
\begin{equation}
R_{\mathrm{trans}} = R |t|^2 P_r \quad \text{and} \quad R_{\mathrm{refl}} = R |r|^2 P_r.
\end{equation}

\section{Two-Excitation Scattering\label{subsectionTwoexci}}

In this section, we apply our general formalism to the specific and important case of two-excitation scattering. We begin in Sec.~\ref{sec:atomic_tmatrix} by formulating the Lippmann-Schwinger equation for the two-excitation atomic T-matrix and presenting its analytic solution for two-level nonlinearities. This result leads to a remarkably simple and separable structure for the on-shell S-matrix, which we detail in Sec.~\ref{SubsectionStructure}. We then explore the consequences of this formalism through several applications. In Sec.~\ref{subsecCritical}, we analyze the S-matrix behavior near the critical energies of the two-dark-spin-wave dispersion, a topic explored in depth in our companion work~\cite{Wang2025UniversalScattering}. In Sec.~\ref{section_dark_states}, we show how the S-matrix provides a framework for identifying multi-excitation dark states. Finally, in Sec.~\ref{sec_cs_two_atomic}, we derive expressions for the two-excitation scattering cross sections and conclude with an illustrative 1D scattering example in Sec.~\ref{sec:1d_example}.

\subsection{The atomic T-matrix}
\label{sec:atomic_tmatrix}

In this subsection, we detail the computation of the two-excitation atomic T-matrix, $\bar{T}^{(2)}$, which is a central quantity for determining the scattering between all channels. We present the general Lippmann-Schwinger equation for $\bar{T}^{(2)}$ and then derive its analytic expression for the case of two-level nonlinearities. This result leads to the remarkably simple expressions for the S-matrix and scattering cross sections presented in the following sections.

To analyze the two-excitation problem, we work in the center-of-mass basis. For a pair of spin waves with lattice momenta $\bm{p}_1$ and $\bm{p}_2$, we define the total momentum $\bm{P}=\bm{p}_1+\bm{p}_2$, which is conserved during scattering, and the relative momentum $\bm{q}=(\bm{p}_1-\bm{p}_2)/2$. The state of such a pair is denoted by $|\bm{P},\bm{q}\rangle$. Due to the indistinguishability of the two bosonic excitations, a state with relative momentum $\bm{q}$ is identical to one with $-\bm{q}$. We therefore define the two-excitation Brillouin zone, $\mathrm{BZ}^{(2)}$, as the restricted range of relative momenta that labels unique states (e.g., $q>0$ in 1D and $q_y>0$ in 2D), as shown by the unshaded regions in Fig.~\ref{fig:D_i}.

The atomic T-matrix elements connecting states with the same total momentum are defined as
\begin{equation}
\bar{T}(\omega,\boldsymbol{P},\boldsymbol{q},\boldsymbol{q}') \equiv \langle \bm{P}, \bm{q}'|\bar{T}^{(2)}(\omega, \boldsymbol{P})|\boldsymbol{P},\bm{q}\rangle,
\end{equation}
where $\bar{T}^{(2)}(\omega, \boldsymbol{P})$ is the T-operator for a fixed total momentum $\bm{P}$, related to the full T-operator via Eq.~\eqref{eqTbarfixP}. These matrix elements are governed by the two-excitation Lippmann-Schwinger equation
\begin{equation}
\begin{split}
\bar{T} (\omega,\boldsymbol{P},\boldsymbol{q},\boldsymbol{q}')
&=U(\boldsymbol{q}-\boldsymbol{q'}) \\
&+\int_{\mathrm{BZ}^{(2)}} d\boldsymbol{q}'' \frac{U(\boldsymbol{q}''-\boldsymbol{q}') }{\omega-\epsilon^{(2)}(\bm{P}, \bm{q}'')}\bar{T} (\omega,\boldsymbol{P},\boldsymbol{q},\boldsymbol{q}'').
\end{split}
\label{eqLS2excitations}
\end{equation}
Here, $\epsilon^{(2)}(\bm{P}, \bm{q})=\epsilon(\boldsymbol{P}/2+\boldsymbol{q})+\epsilon(\boldsymbol{P}/2-\boldsymbol{q})$ is the dispersion of the two non-interacting spin waves. For a generic, momentum-dependent potential $U(\bm{q}-\bm{q}')$, this integral equation must be solved numerically. For two excitations, the full T-matrix is equivalent to its connected part, $\bar{T}_c(\omega)=\bar{T}(\omega)$.

The problem simplifies considerably for the specific case of two-level atoms. The nonlinearity arises from the fact that two excitations cannot occupy the same atom, which corresponds to a hard-core, on-site interaction in real space. In the momentum basis, this translates to a contact interaction where the potential is independent of momentum, i.e., $U(\bm{q}) \to \infty$.

For such a momentum-independent potential, the T-matrix must also be independent of the relative momenta $\bm{q}$ and $\bm{q}'$. This allows for an exact analytical solution to Eq.~\eqref{eqLS2excitations}, which yields
\begin{equation}
 \bar{T}(\omega, \boldsymbol{P}) = - L(\omega, \boldsymbol{P})^{-1}, \label{eqTLrelation}
\end{equation}
where $L(\omega, \bm{P})$ is the local propagator, given by
\begin{equation}
L(\omega, \bm{P}) \equiv \int_{\mathrm{BZ}^{(2)}} d\boldsymbol{q} \frac{1 }{\omega-\epsilon^{(2)}(\bm{P}, \bm{q})}. \label{eqLdefi}
\end{equation}
This remarkably simple, momentum-independent form of the T-matrix is a key result for our analysis, as it greatly simplifies the calculation of the S-matrix and the scattering cross sections for two excitations.

\begin{figure}[t]
    \centering
    \includegraphics[width=1\linewidth]{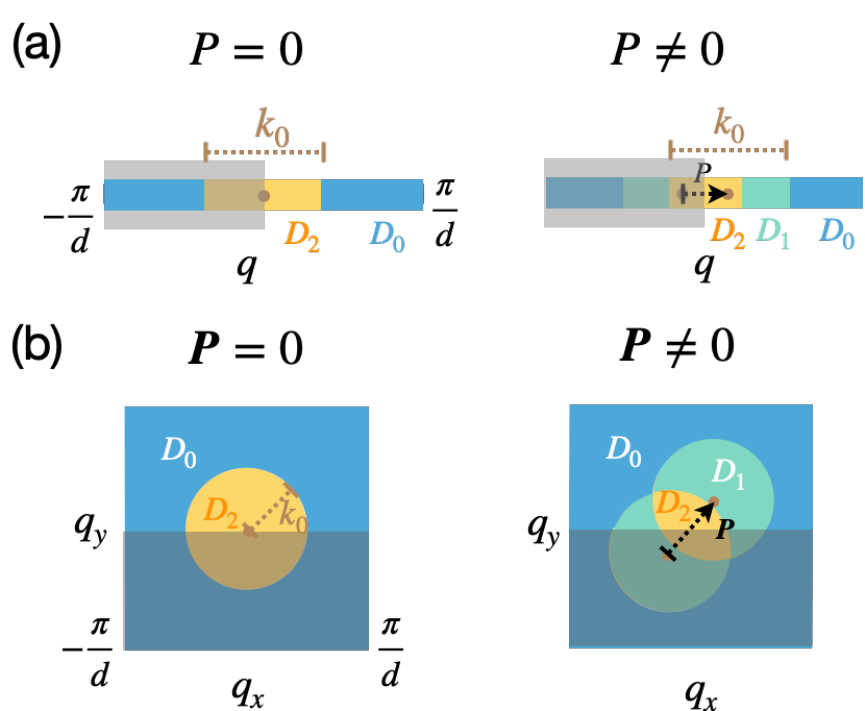}
    \caption{
\textbf{Two-excitation Brillouin zone \(\mathrm{BZ}^{(2)}\).} Partitioning  $\mathrm{BZ}^{(2)}$ (the unshaded region) into the subsets $D_0$, $D_1$, and $D_2$ for the relative momentum $\bm{q}$ of a pair of atomic spin waves with total lattice momentum $\bm{P}$, shown for \textbf{(a)} 1D and \textbf{(b)} 2D atomic arrays.
In each panel, results are shown for $\bm{P}=0$ (left) and $\bm{P} \neq 0$ (right).
The blue regions labeled $D_0$ correspond to domains where both spin waves are dark, the yellow regions ($D_2$) represent both spin waves being bright (within the light cone of radius $k_0$), and the green regions ($D_1$) indicate mixed pairs consisting of one bright and one dark spin wave.
For nonzero $\bm{P}$, the $D_2$ and $D_1$ regions are offset in momentum space by $\bm{P}$, as indicated by the dashed black arrows.
}
    \label{fig:D_i}
\end{figure}

To analyze the contributions from different types of  atomic spin waves, it is useful to decompose the local propagator $L(\omega, \bm{P})$. We begin by partitioning the two-excitation Brillouin zone $\mathrm{BZ}^{(2)}$ into three disjoint subsets $D_{\alpha}(\bm{P})$ for $\alpha \in \{0,1,2 \}$. As illustrated in Fig.~\ref{fig:D_i}, each domain $D_\alpha(\bm{P})$ corresponds to the set of relative momenta $\bm{q}$ for which $\alpha$ spin waves are bright (i.e., inside the light cone) and $2-\alpha$ are dark. This partitioning allows to decompose the local propagator as a sum over components corresponding to these domains, $L(\omega, \bm{P}) = \sum_{\alpha=0}^2 L_\alpha(\omega, \bm{P})$, where each component is defined by integrating over the corresponding domain.

These components have distinct analytical properties. Since the dispersion $\epsilon^{(2)}(\bm{P}, \bm{q})$ is real for two dark spin waves [$\bm{q} \in D_0(\bm{P})$], $L_0(\omega, \bm{P})$ has a branch cut along the real energy axis. In contrast, $L_1(\omega, \bm{P})$ and $L_2(\omega, \bm{P})$ are continuous across the real energy axis because the dispersion for bright spin waves is complex, leading to strictly negative imaginary parts for real energies $\omega=E$.

The imaginary part of $L_0$ has a clear physical meaning: it is directly proportional to the joint density of states for a pair of dark spin waves,
\begin{equation}
\text{Im}[L_0(E\pm i0,\bm{P})] = \mp \pi\rho_0(E,\bm{P}),\label{eqI0_new}
\end{equation}
where $\rho_0(E,\bm{P})=\int_{D_0(\bm{P})} d\bm{q}\ \delta(E-\Delta^{(2)}(\bm{P},\bm{q}))$  and $\Delta^{(2)}(\bm{P},\bm{q})=\Delta(\bm{P}/2+\bm{q})+\Delta(\bm{P}/2-\bm{q})$ is equal to the real part of $\epsilon^{(2)}(\bm{P}, \bm{q})$. Because bright spin waves have complex energies, a direct definition of their density of states is less straightforward. We therefore extend the concept by analogy with the dark spin wave case and define an effective ``off-shell" density of states for pairs containing bright spin waves ($\alpha=1,2$) from the imaginary parts of $L_1$ and $L_2$:
\begin{equation}
\pi\rho_\alpha (E,\bm{P}) \equiv |\text{Im}[L_\alpha(E,\bm{P})]|, \quad \alpha=1,2. \label{eqrhoalpha}
\end{equation}
As we will show, these components of the propagator and their corresponding densities of states are crucial for constructing the multi-channel S-matrix and partial scattering cross sections.

\begin{figure}
    \centering
\includegraphics[width=1\linewidth]{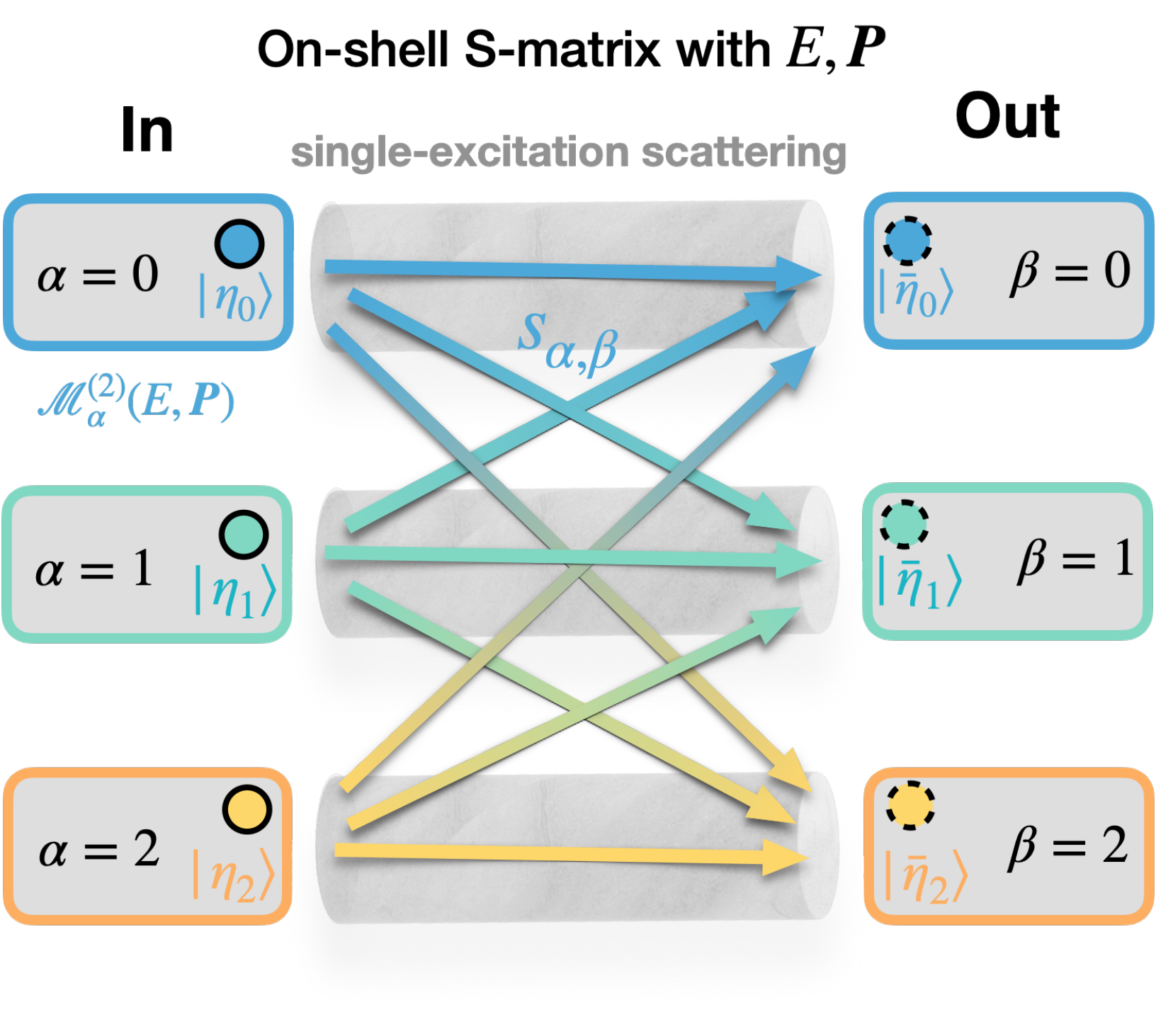}
    \caption{\textbf{Schematic of the simple structure of the on-shell two-excitation S-matrix for two-level nonlinearity.} The scattering process is confined to the on-shell manifold $\mathcal{M}^{(2)}(E,\bm{P})=\sum_{\alpha}\mathcal{M}_{\alpha}^{(2)}(E,\bm{P})$, which is composed of subspaces for each channel $\alpha$ (gray blocks outlined in color: blue $\alpha{=}0$, green $\alpha{=}1$, orange $\alpha{=}2$). For each incoming channel, there is a single special state $|\eta_\alpha\rangle$ (solid circles) that participates in nonlinear scattering. These states scatter into a superposition of the corresponding outgoing states $|\bar{\eta}_\beta\rangle$ (dashed circles) in any open channel $\beta$. This nonlinear process (colored arrows) is described by a $3 \times 3$ unitary matrix with elements $s_{\alpha,\beta}$. All states orthogonal to the $|\eta_\alpha\rangle$ basis undergo only single-excitation scattering (grey tubes), where each excitation scatters independently according to single-excitation dynamics. }
    \label{fig:2Dsummary}
\end{figure}
\subsection{The simple structure of the on-shell S-matrix}
\label{SubsectionStructure}

In this subsection, we illustrate the remarkably simple structure of the two-excitation on-shell S-matrix that emerges as a direct consequence of the momentum-independent atomic T-matrix for two-level nonlinearities. We state the main results here, with detailed derivations provided in  Appendix \ref{sec:smatrix_manifold}.

Due to energy and momentum conservation, the scattering process is restricted to a manifold of states with fixed total energy $E$ and total lattice momentum $\bm{P}$. We denote this on-shell manifold as $\mathcal{M}^{(2)}(E,\bm{P})$, which is the union of the distinct subspaces $\mathcal{M}^{(2)}_{\alpha}(E,\bm{P})$. The on-shell S-matrix, $S^{(2)}(E, \bm{P})$, is the operator that connects the incoming and outgoing states within this manifold.

The key result is that for two-level nonlinearities, the entire two-excitation scattering problem reduces to a simple, separable form. In each channel $\alpha$, there exists a single, special collective state, which we denote $|\eta_\alpha\rangle$, that acts as a ``doorway" for all nonlinear interactions. Any state orthogonal to this special state is unaffected by the nonlinearity and undergoes only linear (single-excitation) scattering.

The special incoming state $|\eta_{\alpha}\rangle$ scatters into a superposition of the corresponding outgoing states $|\bar{\eta}_\beta\rangle$ for all channels $\beta=0,1,2$. The momentum-space wavefunctions  of $|\eta_{\alpha}\rangle$ and $|\bar{\eta}_\beta\rangle$ are complex conjugates of each other.  This reduces the nonlinear part of the scattering to a $3 \times 3$ unitary matrix, with elements $s_{\alpha,\beta}$, that maps the subspace spanned by the orthogonal basis $\{|\eta_\alpha\rangle\}$ to the subspace spanned by $\{|\bar{\eta}_\beta\rangle\}$. The structure of this scattering is illustrated in Fig.~\ref{fig:2Dsummary}. The matrix elements are given by
\begin{equation}
s_{\alpha,\beta}(E,\bm{P})=\delta_{\alpha,\beta} + 2\pi i \frac{\sqrt{\rho_\alpha(E,\bm{P})\rho_\beta(E,\bm{P})}}{L(E+i0,\bm{P})}, \label{eqsalphabeta}
\end{equation}
where $\rho_{\alpha}(E,\bm{P})$ is the effective density of states for a pair of atomic spin waves with $\alpha$ bright excitations, as defined in Eq.~\eqref{eqrhoalpha}. Specifically, for the dark-spin-wave channel, this gives the scattering phase shift $s_{0,0}(E,\bm{P})=L(E-i0, \bm{P}) / L(E+i0,\bm{P})$.

This allows the full on-shell S-matrix to be written as the sum of a nonlinear part acting on the special states and a linear part acting on the orthogonal complement
\begin{equation}
\begin{split}
S^{(2)}(E,\bm{P})&=\sum_{\alpha,\beta=0}^2 s_{\alpha, \beta}|\bar{\eta}_{\beta}\rangle\langle \eta_{\alpha} | \\
&+S^{(2)}_{\mathrm{linear}} (E,\bm{P})\sum_{\alpha=0}^2 \left(\mathbb{1}_{\mathcal{M}^{(2)}_{\alpha}(E,\bm{P})}- |\eta_{\alpha}\rangle \langle \eta_{\alpha}|\right)
\end{split}
\label{eqS_on_shell_simple}
\end{equation}
where  
$S_{\mathrm{linear}}^{(2)}(E,\bm{P})$   denotes the  two-excitation linear-scattering S-matrix [see Eq.\ \eqref{eqSlinear}] acting within the manifold $\mathcal{M}^{(2)}(E,\bm{P})$, and $\mathbb{1}_{\mathcal{M}^{(2)}_{\alpha}(E,\bm{P})}$ is the identity operator on the manifold $\mathcal{M}^{(2)}_{\alpha}(E,\bm{P})$. 

Lastly, we provide the explicit wavefunctions for these special states $|\eta_{\alpha}(E,\bm{P})\rangle$ for the different channels.
For the purely dark-spin-wave channel ($\alpha=0$), the state $|\eta_0\rangle$ has a real-valued wavefunction in the momentum basis, meaning $|\eta_0\rangle = |\bar{\eta}_0\rangle $. It is therefore an eigenstate of the S-matrix with the eigenvalue $s_{0,0}$. Its explicit form is
\begin{equation}
|\eta_{0}(E,\bm{P})\rangle = \frac{1}{\sqrt{\rho_{0}(E,\bm{P})}} \int_{\mathcal{M}^{(2)}_{0}(E,\bm{P})} d\bm{q} \frac{1}{\sqrt{v_g(\bm{P},\bm{q})}} |\bm{P},\bm{q}\rangle,
\end{equation}
where the group velocity $v_g(\bm{P},\bm{q})=\|\nabla_{\bm{q}}\Delta^{(2)}(\bm{P}, \bm{q})\|$.
The nature of the manifold $\mathcal{M}^{(2)}_{0}(E,\bm{P})$ depends on the dimensionality of the system. For a 2D array, it is a 1D equi-energy contour, while for a 1D array, it is a discrete set of points.

For channels involving photons, we denote the basis states by the lattice momenta in the center-of-mass frame ($\bm{P}, \bm{q}$) and the energies of the constituent photons. For the channel with one photon and one dark spin wave ($\alpha=1$), the state $|\eta_{1}(E,\bm{P})\rangle$ is a coherent superposition over all such pairs on the energy manifold. For a given relative momentum $\bm{q} \in D_1(\bm{P})$, one of the momenta $\bm{P}/2 \pm \bm{q}$ corresponds to a dark spin wave (\(\bm{p}_{\mathrm{dark}}\)) and the other to a bright one (\(\bm{p}_{\mathrm{bright}}\)). Energy conservation fixes the photon's energy to be $E_{\mathrm{ph}} = E - \epsilon(\bm{p}_{\mathrm{dark}})$. The state is weighted by the complex conjugate of the atomic amplitude of the bright-spin-wave component of a dressed photon:

\begin{equation}
|\eta_{1}(E,\bm{P})\rangle = \frac{1}{\sqrt{\rho_{1}(E,\bm{P})}} 
\int_{D_1(\bm{P})} d\bm{q} \, a_{\bm{p}_{\mathrm{bright}}}^*(E_{\mathrm{ph}}) |\bm{P},\bm{q},E_{\mathrm{ph}}\rangle,
\end{equation}
where $|\bm{P},\bm{q},E_{\mathrm{ph}}\rangle$ denotes the on-shell state with the appropriate dark and bright-spin-wave components.  As we mentioned, for the outgoing wavefunction $|\bar{\eta}_{1}(E,\bm{P})\rangle$,  $\langle \bm{P},\bm{q},E_{\mathrm{ph}} |\bar{\eta}_{1}(E,\bm{P})\rangle=\langle \bm{P},\bm{q},E_{\mathrm{ph}}  |\eta_{1}(E,\bm{P})\rangle^*$. 

For the two-photon channel ($\alpha=2$), the state $|\eta_{2}(E,\bm{P})\rangle$ is a superposition over all possible two-photon states. For a given relative momentum $\bm{q} \in D_2(\bm{P})$, both spin waves with momenta $\bm{P}/2 \pm \bm{q}$ are bright. The photon energies must sum to the total energy, $E_{\mathrm{ph}} + E'_{\mathrm{ph}} = E$. The state is weighted by the product of the complex conjugates of the two corresponding atomic amplitudes:
\eq{
&|\eta_{2}(E,\bm{P})\rangle = \frac{1}{\sqrt{\rho_{2}(E,\bm{P})}} 
\int_{D_2(\bm{P})} d\bm{q} \int d E_{\mathrm{ph}} \nonumber\\
& a_{\bm{P}/2+\bm{q}}^*(E_{\mathrm{ph}}) a_{\bm{P}/2-\bm{q}}^*(E-E_{\mathrm{ph}}) 
 \times |\bm{P},\bm{q},E_{\mathrm{ph}}, E-E_{\mathrm{ph}}\rangle, \label{eqeta2}
}
where the integral is over all allowed energies for one of the photons. Using Eq.~\eqref{eqintegral_identity}, one can show that the prefactor $1/\sqrt{\rho_{\alpha}(E,\bm{P})}$ in each expression ensures the unity normalization of $|\eta_{\alpha}(E,\bm{P})\rangle$.
As we mentioned, for the output wavefunction $|\bar{\eta}_{2}(E,\bm{P})\rangle$,  $\langle \bm{P},\bm{q},E_{\mathrm{ph}}, E-E_{\mathrm{ph}}|\bar{\eta}_{2}(E,\bm{P})\rangle 
=\langle \bm{P},\bm{q},E_{\mathrm{ph}},E-E_{\mathrm{ph}}|\eta_{2}(E,\bm{P})\rangle^*$.

\subsection{Scattering at critical energies}
\label{subsecCritical}

We now analyze the scattering behavior as the energy $E$ approaches a critical energy $E_{\text{crit}}$. These energies correspond to critical points $\bm{q}_{\text{crit}}$ in the two-dark-spin-wave dispersion where the group velocity vanishes, $\|\nabla_{\bm{q}}\Delta^{(2)}(\bm{P}, \bm{q}_{\text{crit}})\|=0$, a topic explored in depth in Ref.~\cite{Wang2025UniversalScattering}. At these energies, the local propagator $L(E+i0,\bm{P})$ diverges due to the contribution from the dark-spin-waves $L_0(E+i0,\bm{P})$, while the bright-spin-wave contributions remain finite.

This divergence has a profound consequence for the scattering matrix: it forces $s_{\alpha,\beta} = \delta_{\alpha,\beta}$ for all channels involving at least one photon (i.e., where $\alpha$ or $\beta$ is non-zero). This signifies a complete decoupling of the atomic and photonic sectors. In these photon-containing channels, all nonlinear interactions vanish, and each excitation simply undergoes independent, single-excitation scattering. Therefore, nontrivial multi-excitation scattering can \emph{only} occur within the purely atomic dark-spin-wave channel ($\alpha=\beta=0$).

\subsection{Multi-excitation dark states}
\label{section_dark_states}

Multi-excitation dark states are multi-excitation eigenstates of the full photon-atom interacting Hamiltonian with almost only atomic components. Equivalently, they are eigenstates of the non-Hermitian effective atomic Hamiltonian $M^{(n)}$ with eigenvalues whose imaginary parts are zero or near-zero, corresponding to their long lifetimes. Because they are lossless, these dark states are simultaneously right and left eigenstates of $M^{(n)}$. These states are of significant interest for their potential application in multi-photon quantum storage. A key challenge, however, is that a generic combination of $n$ single-excitation dark spin waves is not necessarily an $n$-excitation dark state, as interactions can induce scattering into undesired photonic modes.
Our scattering formalism provides a framework to identify these multi-excitation dark states. Specifically, when the dark-spin-wave S-matrix, $S^{(n)}_{0,0}$, is unitary at a certain energy, it signifies that the dark-spin-wave subspace is decoupled from the photonic channels. Consequently, spin waves within this equi-energy manifold will not scatter into photons and are always dark.

Following this principle, we now demonstrate this analysis within the two-excitation subspace. For a given total lattice momentum $\bm{P}$ and total energy $E_d$, the condition for lossless scattering is that the S-matrix within the dark-spin-wave manifold, $\mathcal{M}^{(2)}_0(E_d,\bm{P})$, must be unitary. This occurs, for example, at the critical energies of the two-spin wave dispersion (see our companion work \cite{Wang2025UniversalScattering}). 
In such cases, any $\bm{q}\in \mathcal{M}^{(2)}_0(E_d,\bm{P})$ defines a two-excitation dark state with the momentum-space representation
\begin{equation}
\begin{split}
&\langle \bm{P}',\bm{q}'|\psi_{\bm{P},\bm{q}}\rangle =   \\
&\delta(\bm{P} - \bm{P}') \bigg[ \delta(\bm{q} - \bm{q}') 
+ \frac{\bar{T}(E_d + i0,\bm{P},\bm{q},\bm{q}')}{E_d + i0 - \epsilon^{(2)}(\bm{P},\bm{q}')} \bigg].
\end{split}
\label{eq2excDark}
\end{equation}

\subsection{Scattering cross section\label{sec_cs_two_atomic}}

We now present explicit expressions for the scattering cross sections in the two-excitation case. To do so, we first specify the notation for the states. A two-excitation state in channel $\alpha$ is described by its momentum configuration, $k^{(2,\alpha)}$:
\begin{equation}
k^{(2,\alpha)} = \begin{cases} \bm{p}_1, \bm{p}_2 \notin \text{LC} & \alpha=0 \\ \bm{p}_1 \in \text{LC}, \bm{p}_2 \notin \text{LC}, \chi & \alpha=1 \\ \bm{p}_1, \bm{p}_2 \in \text{LC}, \chi_1, \chi_2 & \alpha=2 \end{cases}\label{eqk2alpha}
\end{equation}
where $\chi$ and $\chi_{1,2}$ represent the photons' transverse wavenumbers. For any channel $\alpha$, we define the total and relative lattice momenta for the state as $\bm{P} = \bm{p}_1 + \bm{p}_2$ and $\bm{q} = (\bm{p}_1 - \bm{p}_2) / 2$. The group velocity $v_g(k^{(2,\alpha)})$ associated with such a state is given by
\begin{equation}
v_g\big(k^{(2,\alpha)}\big) =
\begin{cases}
\|\nabla_{\bm{q}}\Delta^{(2)}(\bm{P}, \bm{q})\|, & \alpha=0 \\[2ex]
 c & \alpha=1 \\[2ex]
 c\sqrt{2}\sqrt{1-\dfrac{\bm{p}_1 \cdot \bm{p}_2}{k_0^2}}, & \alpha=2
\end{cases} \label{eqvg}
\end{equation}
under the Markov approximation.

Applying the general expression from Eq.~\eqref{eqCrossSectionMain} to an initial state $|k^{(2,\alpha)}\rangle$ gives the partial cross section for scattering into channel $\beta$:
\begin{equation}
\begin{split}
&\sigma^{(2)}_{\alpha, \beta}(k^{(2,\alpha)}) = 4\pi\frac{|a(k^{(2,\alpha)})|^2}{v_g(k^{(2,\alpha)})} \\
&\int_{D_\beta (\bm{P})} d\boldsymbol{q}' \text{Im} \left[ \frac{-1}{E+i0-\epsilon^{(2)}(\bm{P},\boldsymbol{q}')} \right] 
|\bar{T}(E+i0,\bm{P}, \boldsymbol{q},\boldsymbol{q}')|^2.
\end{split}
\label{eqSigmabeta2}
\end{equation}
where the collective atomic amplitude $a(k^{(2,\alpha)})$ is defined in  Eq.~\eqref{eqaknalpha}. 

Heuristically, the integral describes the sum over the scattering probabilities into the atomic-spin-wave subspace $D_\beta(\bm{P})$, with a weight term $\text{Im}[-1/(E+i0-\epsilon^{(2)}(\bm{P},\bm{q}))]$ describing the density of states of the spin wave pair with $\beta$ bright spin waves. When $\beta=0$, $\epsilon^{(2)}(\bm{P}, \bm{q})$ is real, and this weight term is equivalent to a delta function $\pi \delta(E-\epsilon^{(2)}(\bm{P}, \bm{q}))$ in the integral, ensuring energy conservation. When $\beta=1,2$, the weight term accounts for the density of states of the off-shell spin wave pair containing bright spin waves.

In the case of two-level nonlinearity, the atomic T-matrix term $\bar{T}(E,\bm{P})=-L(E,\bm{P})^{-1}$  can be factored out of the integral in  Eq.~\eqref{eqSigmabeta2}.
The remaining integral evaluates to $|\text{Im}[L_\beta(E+i0, \bm{P})]|$, which is directly proportional to the effective density of states $\rho_\beta(E, \bm{P})$ for a pair of spin waves with $\beta$ bright components [see Eqs.~\eqref{eqI0_new} and \eqref{eqrhoalpha}]. This leads to the simplified expression for the partial cross section
\begin{equation}
\sigma^{(2)}_{\alpha,\beta} (k^{(2,\alpha)}) = 4\pi\frac{|a(k^{(2,\alpha)})|^2}{v_g(k^{(2,\alpha)})} \frac{\pi\rho_{\beta}(E, \bm{P})}{|L(E+i0, \bm{P})|^2}. \label{eqsigma2partial}
\end{equation}
The total cross section is then given by
\begin{equation}
\sigma_{\alpha,\mathrm{tot}}^{(2)}(k^{(2,\alpha)}) = 4\pi\frac{|a(k^{(2,\alpha)})|^2}{v_g(k^{(2,\alpha)})} \frac{|\text{Im}[L(E+i0, \bm{P})]|}{|L(E+i0, \bm{P})|^2}. \label{eqsigma_CS}
\end{equation}
Again, when the energy $E$ approaches the critical energies of the two-dark-spin-wave dispersion, $L_0$ diverges, leading to universal behavior in the scattering cross section, which we discuss in our companion paper, Ref.~\cite{Wang2025UniversalScattering}.

\subsection{A 1D array example}
\label{sec:1d_example}

To illustrate the use of our formalism, we consider a simple example: the collision of two dark spin waves in a 1D atomic array polarized along the array direction. Specifically, we take a lattice spacing of~$d = 0.25\lambda_0$, with two incoming dark spin waves characterized by lattice momentum $q$ and $-q$, where $q=\frac{2}{3} \frac{\pi}{d}$ is their relative momentum.  The system's nonlinearity arises from the two-level nature of the atoms.

Since the total momentum $P=0$, conservation of lattice momentum restricts the outgoing state to either a pair of dark spin waves or a pair of photons. In our chosen parameter regime, there exists a unique pair of dark spin waves, $|P=0,q=\frac{2}{3}\frac{\pi}{d}\rangle$, with total energy $E = \epsilon^{(2)}(P=0, q=\frac{2}{3} \frac{\pi}{d})$. Scattering within the same channel ($\alpha = 0$) reproduces this state, yielding a scattering phase $s_{0,0} = -0.6546 - 0.4561i$.

As discussed in Section~\ref{SecCrossSectionDefi}, the partial scattering cross sections in this scenario are dimensionless. Using Eq.~\eqref{eqsigma2partial}, we find $\sigma^{(2)}_{0,0} =|s_{0,0}-1|^2= 2.946$ and $\sigma^{(2)}_{0,2} = 0.364$. The former quantity corresponds to the squared amplitude of the scattered wave in output channel $\beta=0$ and is upper-bounded by $4$. The latter represents the probability of scattering into the two-photon output state $|\bar{\eta}_2\rangle$ as defined in Eq.~\eqref{eqeta2}.

We parametrize the outgoing two-photon wavefunction as \(\bar{\eta}_2(q, \Delta_{\mathrm{ph}}) \equiv \langle P=0, q, E/2+\Delta_{\mathrm{ph}}, E/2-\Delta_{\mathrm{ph}} | \eta_2(E, P=0) \rangle\), where \(\pm q\) and \(E/2 \pm \Delta_{\mathrm{ph}}\) are the lattice momenta and energies of the two outgoing photons, respectively. The total energy of the incoming state is \(E=2.21\Gamma_0\). Due to the indistinguishability of the photons and the symmetries of the $P=0$ state, the wavefunction satisfies \(\bar{\eta}_2(q, \Delta_{\mathrm{ph}}) = \bar{\eta}_2(\pm q, \pm \Delta_{\mathrm{ph}})\).

Figure~\ref{fig:two_photon_output} shows the phase and relative modulus of this wavefunction, normalized to its value at the reference point \((q=0, \Delta_{\mathrm{ph}}=0)\). The plots cover the domain \(q \in [0, 0.5\pi/d)\), which corresponds to the region \(D_2(P=0)\) where both constituent spin waves are bright. Two prominent features are revealed. First, a singularity appears at \(\Delta_{\mathrm{ph}} \approx 1.29\Gamma_0\) and the domain boundary \(q=0.5\pi/d\). This occurs when the energy of one of the outgoing photons becomes resonant with a spin-wave excitation at the edge of the light cone, where radiative decay vanishes. Second, the modulus of the wavefunction is maximized relative to the reference point at \(q=0\) and \(\Delta_{\mathrm{ph}} \approx 2.27\Gamma_0\).

\begin{figure}
    \centering
\includegraphics[width=1\linewidth]{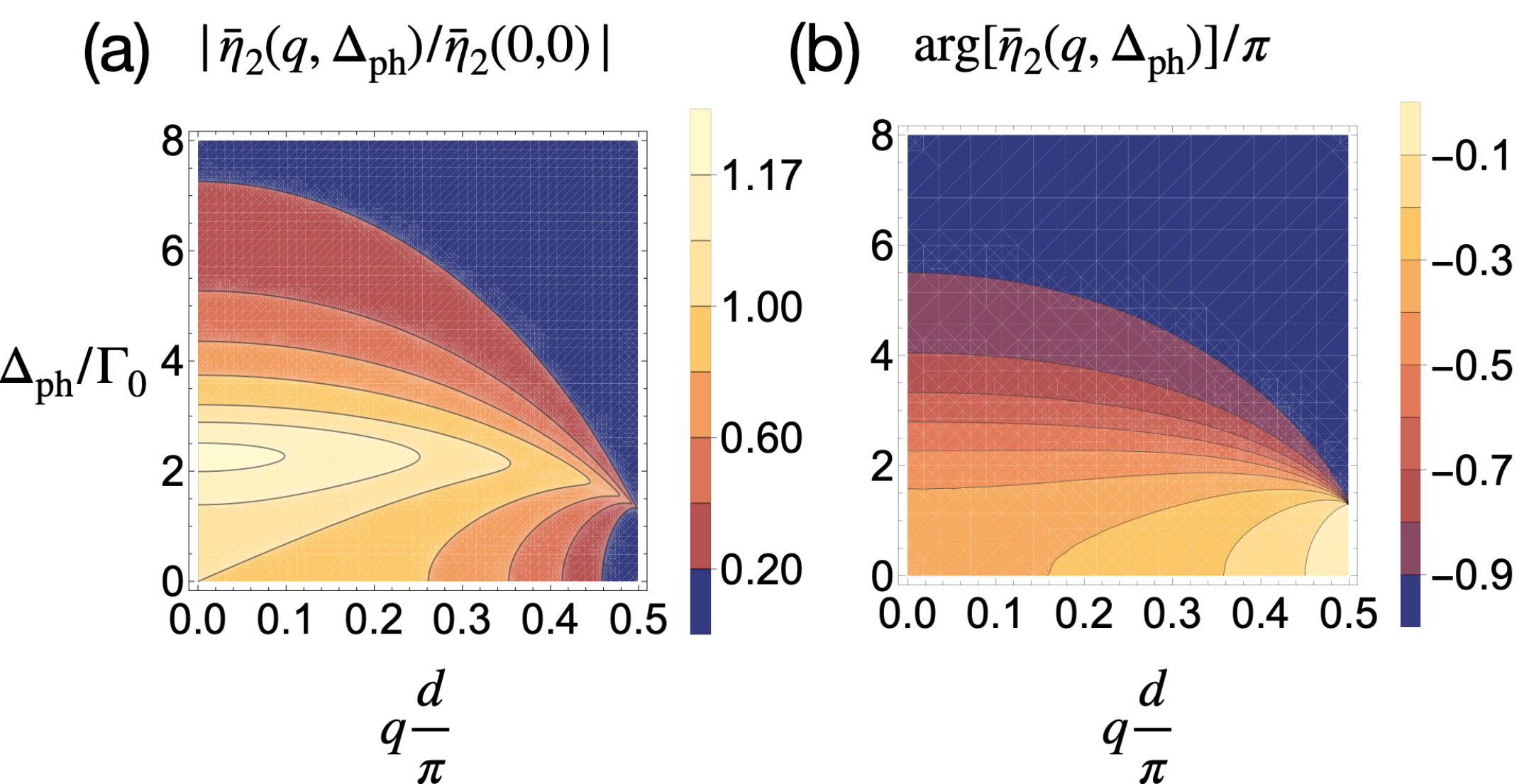}
    \caption{
Wavefunction $\bar{\eta}_2(q,\Delta_{\mathrm{ph}})$ of the two-photon outgoing state resulting from the collision of two dark spin waves with momenta $\pm\frac{2}{3}\frac{\pi}{d}$ in a 1D array with lattice constant $d = 0.25 \lambda_0$ and atoms polarized along the array. The photon lattice momenta and energies are $\pm q$  and $E/2\pm \Delta_{\mathrm{ph}}$, respectively, where $E$ is the total energy of the incoming state. Panel \textbf{(a)} shows the relative amplitude $|\bar{\eta}_2(q,\Delta_{\mathrm{ph}})/\bar{\eta}_2(0,0)|$, referenced to its value at $(q=0,\Delta_{\mathrm{ph}}=0)$, while panel \textbf{(b)} displays the phase of $\bar{\eta}_2(q,\Delta_{\mathrm{ph}})$.
}
    \label{fig:two_photon_output}
\end{figure}

\section{Conclusion}

In this work, we have developed a comprehensive, first-principles scattering theory for multi-excitation processes in subwavelength atomic arrays. The main result is a powerful analytical reduction of this complex multi-channel problem. We have shown that the two fundamental quantities of scattering theory, the S-matrix (which encodes all transition amplitudes) and the scattering cross sections (which quantifies scattering efficiency), are both fully determined by the T-matrix defined purely within the atomic-spin-wave subspace.

As a key application of this framework, we have explicitly solved the two-excitation problem for systems with two-level nonlinearities, deriving the general, analytic form of the S-matrix and scattering cross sections for 1D and 2D arrays. This  solution provides the rigorous theoretical foundation for the universal scattering phenomena in 2D arrays reported in our companion Letter \cite{Wang2025UniversalScattering}. The practical implementation of the formalism is then demonstrated through an illustrative calculation for a 1D system.
This analytical toolkit opens several avenues for future research. For quantum information science, the ability to calculate exact scattering observables can guide the development of novel protocols for the deterministic generation of multi-photon entanglement, efficient quantum storage in dark spin waves, and the creation of hybrid photon-spin-wave entanglement. From a fundamental perspective, it would be fascinating to generalize our methods to study the formation and interaction of multi-photon and multi-spin-wave bound states~\cite{PhysRevLett.132.163602}, particularly in systems with more general atomic nonlinearities. 

We note that while our formalism is exact, obtaining analytical solutions for problems beyond two excitations presents a significant computational challenge, as the complexity of the T-matrix grows rapidly with excitation number. Ultimately, by providing a complete and rigorous method for computing scattering observables from first principles, this work lays a solid foundation for future explorations of the rich few- and many-body physics emerging in these versatile quantum optical platforms.

\section*{Acknowledgement}
We thank Alec Douglas, Simon Hollerith, Sayed Ali Akbar Ghorashi for discussions.  Y.W. and S.Y. acknowledge support from the NSF through CUA PFC (PHY-2317134), PHY-2207972, and the Q-SEnSE QLCI (OMA-2016244). O.R.-B. acknowledges support from Fundación Mauricio y Carlota Botton and from Fundació Bancaria “la Caixa” (LCF/BQ/AA18/11680093). V. W. acknowledges support from the NSF through Grant No. PHYS-2409630 for early-stage investigations of the scattering formalism, and support from the U.S. Department of Energy, Office of Science, Office of Basic Energy Sciences Energy Frontier Research Centers program under Award Number DE-SC0025620 for the subsequent analysis of the dynamics near critical points.
\appendix

\section{The system details}
\label{app:system_details}
This appendix provides the foundational definitions for the Hamiltonians and operators used in the main text. Section~\ref{app:full_hamiltonian} presents the full real-space Hamiltonian and defines the collective atomic spin-wave modes $b^\dagger_{\bm{p}}$. Section~\ref{app:photon_modes} then provides the explicit construction of the effective photon modes $C^\dagger_{\bm{p}}(\chi)$ from the free-space modes. 

\subsection{Real-space Hamiltonian and momentum-space representation}
\label{app:full_hamiltonian}
We consider an array of two-level atoms with aligned dipole moments $\bm{\mu}$, coupled to the free-space vacuum.  The full Hamiltonian is $H_{\text{tot}} = H_{\text{quad}} + U$, where we work in units of $\hbar=1$:
\begin{subequations}
\begin{align}
& H_{\text{quad}} = \sum_{i=1}^N \omega_{eg} b^\dagger_{i} b_{i}
  + \sum_{\Lambda=\pm} \int d\boldsymbol{k}\ c\|\boldsymbol{k}\| \mathcal{E}_{\Lambda}^{\dagger}(\boldsymbol{k}) \mathcal{E}_{\Lambda}(\boldsymbol{k}) \nonumber\\
&\quad + \sum_{i, \Lambda} \int d\boldsymbol{k}\ [g_{\bm{k},\Lambda}\exp(i\bm{k}\cdot\bm{r}_i) \mathcal{E}_{\Lambda}(\boldsymbol{k}) b^{\dagger}_{i} + \text{h.c.}], \\
& \quad \quad    U  = \frac{1}{2}\sum_{i \neq j} \tilde{U}(\bm{r}_i - \bm{r}_j) b^\dagger_i b^\dagger_j b_i b_j.
\end{align}
\end{subequations}
Here, $b_i^\dagger$ are creation operators for an atomic excitation at site $i$. These excitations are treated as hard-core bosons, with the operators satisfying the canonical commutation relations $[b_i, b^\dagger_j]=\delta_{ij}$.

Here, $b_i^\dagger$ are hard-core boson creation operators for an excitation at site $i$. $[b_i, b^\dagger_j]=\delta_{ij}$. We consider atoms in regular 1D or 2D square lattices with lattice constant $d$, in the thermodynamic limit ($N \to \infty$). The operator $\mathcal{E}_\pm^\dagger(\boldsymbol{k})$ creates a plane-wave photon with momentum $\bm{k}$ and helicity $\Lambda=\pm$. The atom-photon coupling is $g_{\bm{k},\Lambda}= \bm{\mu}\cdot \bm{\epsilon}_{\bm{k},\Lambda}\sqrt{\frac{ \omega_{eg}}{2\epsilon_0 v}}$, where $ \bm{\epsilon}_{\bm{k},\Lambda}$ is the photon polarization vector and $v$ is the quantization volume.

Leveraging the translational symmetry of the array, we define the creation operator for a collective atomic spin wave with lattice momentum $\bm{p}$ as

\begin{equation}
b^\dagger_{\boldsymbol{p}} \equiv \frac{1}{\sqrt{N}} \sum_{i=1}^{N} \exp(i\boldsymbol{p} \cdot \boldsymbol{r}_{\| , i}) b_{i}^\dagger,
\end{equation}
where $\bm{r}_{\|, i}$ is the position vector of the $i$-th atom within the array's dimension.

\subsection{Definition of effective photon modes}
\label{app:photon_modes}

Here we provide the explicit expression of $C^\dagger_{\bm{p}}(\chi)$ for 1D and 2D regular atomic arrays, applicable to different atomic polarizations.

\subsubsection{2D atomic arrays}
\label{subsec2DC}

For a 2D atomic array, the effective photon mode $C_{\bm{p}}^\dagger(\chi)$ is constructed from the subset of free-space photon modes that share the same lattice momentum $\bm{p}$. This construction combines contributions from both helicities ($\Lambda=\pm$) and from photons propagating away from the array in the $+z$ and $-z$ directions. The general form is a linear superposition weighted by the coupling strengths $g_{\bm{k},\Lambda}$:
\begin{equation}
 C_{\bm{p}}^\dagger (\chi)=\mathcal{N}\sum_{\Lambda=\pm }\left[ g_{\bm{k},\Lambda}   \mathcal{E}_{\Lambda}^{\dagger} (\bm{k}) +g_{\bar{\bm{k}},\Lambda}  
  \mathcal{E}_{\Lambda}^{\dagger} (\bm{\bar{k}})\right],\label{eqCpdefi}
\end{equation}
where $\bm{k}=(\bm{p}, \chi)^T$ and $\bar{\bm{k}}=(\bm{p}, -\chi)^T$. The normalization factor $\mathcal{N}$ is chosen to ensure that $C_{\bm{p}}^\dagger (\chi)$ satisfies canonical bosonic commutation relations.

The simplification of this expression for different atomic polarizations arises from the fundamental relation $g_{\bm{k},\Lambda} \propto \bm{\mu} \cdot  \bm{\epsilon}_{\bm{k},\Lambda}$. To perform this calculation, we express the photon polarization vectors $ \bm{\epsilon}_{\bm{k},\Lambda}$ in the spherical basis $\{ \mathbf{e}_0,  \mathbf{e}_{\pm 1}\}$:
\begin{equation}
 \mathbf{e}_0 = \mathbf{e}_z, \quad  \mathbf{e}_{\pm 1} = \mp \frac{1}{\sqrt{2}}(\mathbf{e}_x \pm i\mathbf{e}_y),
\end{equation}
where $\mathbf{e}_x, \mathbf{e}_y, \mathbf{e}_z$ are the Cartesian unit vectors (The 2D array is in the $x,y$ plane). The expansions are given by
\begin{subequations}
\begin{align}
 \bm{\epsilon}_{\boldsymbol{k}, +} &= \frac{1}{\sqrt{2}}\sin(\theta) \mathbf{e}_{0} + \exp(-i\phi)\cos^2(\theta/2) \mathbf{e}_+ \nonumber\\
&\quad + \exp(i\phi)\sin^2(\theta/2) \mathbf{e}_-, \\
 \boldsymbol{\epsilon}_{\boldsymbol{k}, -} &= -\frac{1}{\sqrt{2}}\sin(\theta) \mathbf{e}_{0} + \exp(-i\phi)\sin^2(\theta/2) \mathbf{e}_+ \nonumber\\
&\quad + \exp(i\phi)\cos^2(\theta/2) \mathbf{e}_-.
\end{align}
\label{eqekpm}
\end{subequations}
where the vectors for $\bar{\bm{k}}$ are found by the substitution $\theta \to \pi-\theta$.

When the atoms are perpendicularly polarized, i.e. $\bm{\mu}$ is parallel to  $ \mathbf{e}_0$, the dot product $\bm{\mu} \cdot  \bm{\epsilon}_{\bm{k},\Lambda}$ selects the coefficient of the $ \mathbf{e}_0$ term in the expansions of $ \bm{\epsilon}_{\bm{k},\Lambda}$. This directly leads to the symmetry relations $g_{\bm{k},+} = -g_{\bm{k},-}$ and $g_{\bar{\bm{k}},+} = -g_{\bar{\bm{k}},-}$. Substituting these into Eq.~\eqref{eqCpdefi} and normalizing yields
\begin{equation}
C^\dagger_{\boldsymbol{p}}(\chi) = \frac{1}{2}\big[ \mathcal{E}_{+}^\dagger(\boldsymbol{k}) - \mathcal{E}_{-}^\dagger(\boldsymbol{k}) + \mathcal{E}_+^\dagger(\boldsymbol{\bar{k}}) - \mathcal{E}_-^\dagger(\boldsymbol{\bar{k}}) \big].
\end{equation}

When the atoms are circularly polarized, i.e. $\bm{\mu}$ is parallel to $ \mathbf{e}_{\pm}$,
\begin{align}
C^\dagger_{\bm{p}}(\chi) &= \frac{1}{\sqrt{2}}\mathcal{N}\big[\cos^2(\theta/2)\mathcal{E}_\pm^\dagger(\bm{k}) + \sin^2(\theta/2)\mathcal{E}_\mp^\dagger(\bm{k}) \nonumber\\
&\quad + \sin^2(\theta/2)\mathcal{E}_\pm^\dagger(\bm{\bar{k}}) + \cos^2(\theta/2)\mathcal{E}_\mp^\dagger(\bm{\bar{k}})\big],
\end{align}
where the normalization constant is $\mathcal{N} = (1 - \frac{1}{2}\sin^2\theta)^{-1/2}$. 
\subsubsection{1D Atomic Arrays} \label{subsec1DC}

For a 1D array of atoms placed along the $z$-axis, the collective photonic mode $C_{p}^\dagger(\chi)$ is given by an integral over plane-wave modes:
\eq{
 C_{p}^\dagger (\chi) = \mathcal{N}_{1D} \sum_{\Lambda=\pm} \int_0^{2\pi} d\theta\  g_{\bm{k},\Lambda} \mathcal{E}_{\Lambda}^{\dagger} (\bm{k}), \label{eqCdefi1D}
}
where the wavevector is $\bm{k}=[\chi\cos(\theta), \chi\sin(\theta), p]^T$ and $\mathcal{N}_{1D}$ is a normalization constant. $\theta$ is the azimuthal angle in the $xy$-plane, while the polar angle is fixed by the momentum components, defined as $\theta_k \equiv \arctan(\chi/p)$. Due to the system's cylindrical symmetry, it is advantageous to re-express this mode in the basis of cylindrical Bessel photons (also known as ``twisted photons") \cite{knyazev2018beams, ababekri2024vortex}.

These basis states are described by the vector wavefunctions $\bm{A}_{\chi, m, p, \Lambda}(\bm{r})$, where $\bm{r}$ is the position vector expressed in cylindrical coordinates as $(r, \phi_r, z)$. These wavefunctions represent photons with a well-defined momentum $p$ along the $z$-axis, energy $\sqrt{\chi^2+p^2}$, total angular momentum projection $m$ along $z$, and helicity $\Lambda$

\eq{
&\bm{A}_{\chi, m, p, \Lambda}(\bm{r}) = \nonumber\\
&\sum_{\sigma=0,\pm 1} i^{-\sigma} d_{\sigma ,\Lambda}(\theta_k) J_{m-\sigma}(\chi r)\exp[i(m-\sigma)\phi_r + ip z]  \mathbf{e}_{\sigma},
}
The total angular momentum projection $m$ is the sum of the spin ($m_s = \sigma$) and orbital ($m_l = m - \sigma$) contributions. The coefficients $d_{\sigma, \Lambda}(\theta_k)$ are the small Wigner d-matrices, which for helicity $\Lambda = \pm 1$ are given by:
\begin{subequations}
\eq{
&d_{\Lambda, \Lambda}(\theta_k) =\cos^2(\theta_k/2), \\
&d_{-\Lambda,\Lambda}(\theta_k)=\sin^2(\theta_k/2), \\
&d_{0,\Lambda}(\theta_k)=\frac{\Lambda}{\sqrt{2}}\sin(\theta_k).
}
\end{subequations}
The corresponding creation and annihilation operators, $\bm{A}_{\chi, m, p, \Lambda}^\dagger$ and $\bm{A}_{\chi, m, p, \Lambda}$, satisfy the bosonic commutation relations for a continuous orthogonal basis \cite{knyazev2018beams}:
\eq{
[\bm{A}_{\chi, m, p, \Lambda}, \bm{A}_{\chi', m', p', \Lambda'}^\dagger] = \frac{4\pi^2}{\chi} \delta(\chi - \chi') \delta(p - p') \delta_{m,m'} \delta_{\Lambda,\Lambda'}. \label{eq:A_commutation}
}

A crucial simplification occurs because the atoms are located at $r=0$. The Bessel function $J_{n}(0)$ is non-zero only for $n=0$, meaning the atoms couple exclusively to photon modes where $m - \sigma = 0$, which implies $m=\sigma$. This restricts the interaction to twisted photons with total angular momentum projection $m \in \{0, \pm 1\}$. All modes with $|m| \ge 2$ are completely decoupled from the 1D atomic array.

Consequently, the integral in Eq.~\eqref{eqCdefi1D} reduces to a discrete sum over these three interacting channels:
\eq{
 C_{p}^\dagger (\chi) = \mathcal{N}'_{1D} \sum_{\Lambda=\pm} \sum_{m=0, \pm 1} (\bm{\mu} \cdot \bm{A}_{\chi, m, p,\Lambda}^{\dagger}). \label{eq:C_p_sum_twisted}
}
where $\mathcal{N}'_{1D}$ is the normalization constant. 
If the atom polarizations are transverse to the array ($\bm{\mu} \parallel  \mathbf{e}_{\pm}$),
\eq{
 C_{p}^\dagger (\chi) = \mathcal{N}_{\parallel }\left[\cos^2(\theta_k/2)\bm{A}^\dagger_{\chi, \pm 1, p, \pm}
+\sin^2(\theta_k/2)\bm{A}^\dagger_{\chi, \pm 1, p, \mp}\right],
}
where $\mathcal{N}_{\parallel }=\left(1-\frac{1}{2}\sin^2(\theta_k)\right)^{-1/2}$. If the atom polarizations are parallel to the array ($\bm{\mu} \parallel  \mathbf{e}_{0}$),
\eq{
 C_{p}^\dagger (\chi) = \frac{1}{\sqrt{2}}(\bm{A}^\dagger_{\chi, 0, p,+} - \bm{A}^\dagger_{\chi, 0, p,- }).
}

\section{Single-photon scattering eigenstates}
\label{app:single_photon_states}

In this appendix, we provide the explicit wavefunctions of the single-excitation eigenstates (dressed photons) of the subspace Hamiltonian $H_{\bm{p}}$. The creation operator for an eigenstate with incoming photon momentum $\chi$ is a superposition of the atomic and photonic modes:
\begin{equation}
\psi^\dagger_{\bm{p}}(\chi) =\int_{-\infty}^{+\infty}d\chi'  \psi_{\bm{p},\chi}(\chi') C^\dagger_{\bm{p}}(\chi') + a_{\bm{p}}(E_{\bm{p}}(\chi))b^\dagger_{\boldsymbol{p}},
\label{eqdressedPh_app}
\end{equation}
where $a_{\bm{p}}(E)$ is the atomic amplitude from Eq.~\eqref{eqepchi}, and the photonic part of the eigenstate, $\psi_{\bm{p},\chi}(\chi')$, is given by 
\begin{equation}
\psi_{\bm{p},\chi}( \chi' )=\delta(\chi-\chi') + a_{\bm{p}}(E_{\bm{p}}(\chi))\frac{g_{\bm{p}}}{E_{\bm{p}}(\chi)+i0-E_{\bm{p}}(\chi')}.
\label{eqPsi_p_chi_app}
\end{equation}

These dressed-photon eigenstates form a complete and orthonormal basis for the single-excitation subspace with lattice momentum $\bm{p}$, satisfying the bosonic commutation (orthonormality) relation
\begin{equation}
[\psi_{\bm{p}}(\chi), \psi^\dagger_{\bm{p}'}(\chi')] = \delta(\bm{p}-\bm{p}')\delta(\chi - \chi'),
\end{equation}
and the completeness relation
\begin{equation}
\int d\chi |\psi_{\bm{p}}(\chi)\rangle\langle\psi_{\bm{p}}(\chi)| = \mathbb{1}_{\bm{p}},
\end{equation}
where $\mathbb{1}_{\bm{p}}$ is the identity operator in this subspace. Consequently, the Hamiltonian $H_{\bm{p}}$ is diagonal in this basis:
\begin{equation}
H_{\bm{p}}=\int d\chi \, E_{\bm{p}}(\chi)\psi^{\dagger}_{\bm{p}}(\chi)\psi _{\bm{p}}(\chi).
\label{eqHpdiagonal_app}
\end{equation}

To provide an intuitive physical picture of this scattering process, we map the system to an effective 1D real-space channel by performing a Fourier transform on the momentum-space wavefunction $\psi_{\bm{p},\chi}(\chi')$. Since the physical photon modes exist only for $\chi' > 0$, we formally extend the domain to $(-\infty, +\infty)$ by including fictitious modes for $\chi' < 0$. This mathematical step is justified as these modes are far off-resonant from the atomic transition and do not participate in the dynamics. Performing the Fourier transform yields the real-space wavefunction
\eq{
\psi_{\bm{p},\chi}(r) &= \frac{1}{\sqrt{2\pi}} \int_{-\infty}^{+\infty} d\chi' \, e^{ir\chi'}  
\psi_{\bm{p},\chi}(\chi') \\
&= \frac{1}{\sqrt{2\pi}}\left[e^{i\chi r}\theta(-r)+t_{\bm{p}}(E_{\bm{p}}(\chi))e^{i\chi r}\theta(r)\right].
}
This result has a clear physical interpretation in the effective 1D  channel: an incoming plane wave from $r<0$ scatters off the emitter at $r=0$, continuing as a transmitted wave for $r>0$ with an amplitude given by the transmission coefficient $t_{\bm{p}}(E)$ [Eq.~\eqref{eqt_pchi}].

\section{Derivation of the key integration identity\label{appenKeyIdentity}}

In this appendix, we derive the identity, Eq.\ \eqref{eqintegral_identity}, used to analytically evaluate the integrals over orthogonal photon wavenumbers in the partial scattering cross section:
\begin{align}
&\pi \int d\chi_1 \cdots d\chi_\beta\; \delta\left( E - \sum_{j=1}^\beta E_{\bm{p}_j}(\chi_j) \right) \prod_{j=1}^\beta |a_{\bm{p}_j}(E_{\bm{p}_j}(\chi_j))|^2 \nonumber\\
&= -\mathrm{Im} \frac{1}{E + i0 - \sum_{j=1}^\beta \epsilon(\bm{p}_j)},
\end{align}
where  $a_{\bm{p}}(E)$ is the atomic amplitude excited by a single photon with energy $E$, $E_{\bm{p}}(\chi)$ is photon dispersion and $\epsilon(\bm{p})$ is the dispersion relation of an atomic spin wave.

\subsection*{Case $\beta = 1$}

We first consider the case of a single photon. Using the resolution of the identity for the set of dressed photon eigenstates $|\psi_{\bm{p}}(\chi)\rangle$,
\begin{equation}
|a_{\bm{p}}(E_{\bm{p}}(\chi))|^2 = \langle 0|b_{\bm{p}}|\psi_{\bm{p}}(\chi)\rangle\, \langle \psi_{\bm{p}}(\chi)|b^\dagger_{\bm{p}}|0\rangle,
\end{equation}
and the relation
\begin{equation} 
\pi \delta(E - E_{\bm{p}}(\chi) ) = -\mathrm{Im} \left[ \frac{1}{E + i0 -E_{\bm{p}}(\chi) } \right],
\end{equation}
we find 
\begin{align}
&\pi \int d\chi\, \delta(E - E_{\bm{p}}(\chi) )\, |a_{\bm{p}}(E_{\bm{p}}(\chi))|^2 \nonumber\\
&= -\mathrm{Im} \left[ \int d\chi\, \langle 0|b_{\bm{p}}|\psi_{\bm{p}}(\chi)\rangle\, \frac{1}{E + i0 - E_{\bm{p}}(\chi) } \langle \psi_{\bm{p}}(\chi)|b^\dagger_{\bm{p}}|0\rangle \right].
\end{align} 
By invoking the mode decomposition [Eq.~\eqref{eqHpdiagonal_app}] of the single-excitation Hamiltonian \(H_{\bm{p}}\) [Eq.~\eqref{eq:H_p_LC_intro}], this integral yields
\begin{equation}
-\mathrm{Im} \left[ \langle 0|b_{\bm{p}}\, \frac{1}{E + i0 - H_{\bm{p}}} b^\dagger_{\bm{p}}|0\rangle\right].
\end{equation} Under the Markov approximation, this Green's function reduces to
\begin{equation}
-\mathrm{Im} \left[ \frac{1}{E + i0 - \epsilon(\bm{p})}\right].
\end{equation}
This establishes the identity for $\beta = 1$.

\subsection*{Case $\beta \geq 1$}

For arbitrary $\beta$, the product structure of the final state means each integral can be evaluated separately, so the result generalizes by repeated application of the above:
\begin{align}
&2\pi \int d\chi_1 \cdots d\chi_\beta \; \delta\left(E - \sum_{j=1}^\beta E_{\bm{p}_j}(\chi_j)\right) \prod_{j=1}^\beta |a_{\bm{p}_j}(E_{\bm{p}_j}(\chi_j))|^2 \nonumber \\
&= -\mathrm{Im} \frac{1}{E + i0 - \sum_{j=1}^\beta \epsilon(\bm{p}_j)}.
\end{align}
This completes the derivation of the key integration identity used in the calculation of the partial scattering cross section.

\section{Derivation of coherent state transmission}
\label{app:coherent_transmission}

In this appendix, we provide a full derivation for the transmission of a weak coherent beam incident on a 2D atomic array, focusing on the effect of nonlinear interactions. Specifically, we derive the relationship between the beam's survival probability $P_r$ at the output and the multi-excitation cross sections, leading to Eq.~\eqref{eqPsurvival}. 

Consider a long, uniform light pulse in a coherent state $|\alpha_{\mathrm{c}}\rangle$ incident from one side onto the 2D atomic array. We model the pulse as having a duration $T$ and a cross-sectional area $A$, such that the annihilation operator for this mode is $a$. The coherent state is given by
\eq{
|\alpha_{\mathrm{c}}\rangle_{\mathrm{in}} = \exp\left(-\frac{|\alpha_{\mathrm{c}}|^2}{2}\right) \sum_{n=0}^{\infty} \frac{(\alpha_{\mathrm{c}})^n}{\sqrt{n!}} |n\rangle,
}
where $|n\rangle $ is a Fock state with $n$ photons. The incident photon flux (photons per unit time per unit area) is $R = \frac{|\alpha_{\mathrm{c}}|^2}{A T}$.

The nonlinear scattering process causes photons to be removed from the initial coherent mode by multi-photon scattering events. To quantify this loss, we focus on the average number of photons lost from the coherent beam due to nonlinear scattering, denoted as $N_{\mathrm{loss}}$. This average number of lost photons is calculated by summing over all possible initial photon numbers and all possible nonlinear loss events:
\eq{
N_{\mathrm{loss}} = \sum_{n=2}^{\infty} \left( \exp(-|\alpha_{\mathrm{c}}|^2) \frac{|\alpha_{\mathrm{c}}|^{2n}}{n!} \right)  \sum_{m=2}^{n} m \binom{n}{m} p^{(m)} .
}
Here, the first term in parentheses represents the Poisson probability of having $n$ photons initially. The inner sum represents the total number of photons removed from an $n$-photon state due to connected nonlinear scattering: $m$ is the number of photons lost in an $m$-photon event, $\binom{n}{m}$ is the number of ways to choose $m$ photons out of $n$ to scatter, and $p^{(m)}$ is the probability of a single, connected $m$-photon scattering event for a specific group of $m$ photons:
\eq{
p^{(m)} = \frac{\sigma^{(m)}_{m,\mathrm{tot}}}{(A T)^{m-1}}.
}
Here, $\sigma^{(m)}_{m,\mathrm{tot}}$ is the total $m$-photon scattering cross section, $A$ is the cross-sectional area of the beam, and $T$ is the pulse duration.

Performing the summations and substitutions, this expression for $N_{\mathrm{loss}}$ simplifies to
\eq{
N_{\mathrm{loss}} = \sum_{m=2}^{\infty} \frac{p^{(m)}}{(m-1)!} |\alpha_{\mathrm{c}}|^{2m}.
}

The average number of photons \emph{remaining} in the transmitted coherent mode, $N_{\mathrm{out}}$, is simply the initial average number of photons minus those lost: $N_{\mathrm{out}} = |\alpha_{\mathrm{c}}|^2 - N_{\mathrm{loss}}$. The nonlinear survival probability $P_r$ is then defined as the ratio of average number of photons in the output beam to the initial average number of photons:
\eq{
P_r = 1 - \frac{N_{\mathrm{loss}}}{|\alpha_{\mathrm{c}}|^2} = 1 - \sum_{m=2}^{\infty} \frac{\sigma^{(m)}_{m,\mathrm{tot}}}{(m-1)!} R^{m-1}.
}
This completes the quantum-optics derivation of the nonlinear survival probability $P_r$ in Eq.~\eqref{eqPsurvival}.

\section{Derivation of the two-excitation on-shell S-matrix}
\label{sec:smatrix_manifold}

In this appendix, we derive Eq.~\eqref{eqS_on_shell_simple} from the main text, which gives the simplified form of the on-shell S-matrix in the two-excitation subspace for a system with two-level nonlinearity.
We denote by $S^{(2)}_{\alpha, \beta}(E, \bm{P})$ the on-shell S-matrix component connecting  incoming states in channel $\alpha$ to  outgoing states in channel $\beta$. The full on-shell S-matrix is thus
\eq{
S^{(2)}(E,\bm{P}) = \sum_{\alpha,\beta=0}^2 S_{\alpha,\beta}(E,\bm{P}).
}

Given that $\bar{T}(E+i0, \bm{P}) = -L(E+i0,\bm{P})^{-1}$ for two-level nonlinearities, projecting the two-excitation version of Eq.~\eqref{eqSgen} onto the energy-momentum manifold yields
\begin{subequations}
\begin{align}
S^{(2)}_{\alpha, \beta}(E, \bm{P}) & = \delta_{\alpha,\beta}
S_{\mathrm{linear},\alpha}^{(2)}(E,\bm{P}) \nonumber \\
&\quad + \tau_{\alpha,\beta}(E,\bm{P})\, |\bar{\eta}_{\beta}(E,\bm{P})\rangle \langle \eta_{\alpha}(E,\bm{P})|.
\label{eq:StwobodyManifold_restructured}
\end{align}
The first term describes linear scattering, where $S_{\mathrm{linear},\alpha}^{(2)}(E,\bm{P})$ is the restriction of the linear S-matrix $S_{\mathrm{linear}}^{(2)}(E,\bm{P})$  to channel  $\alpha$. 
The second term of Eq.~\eqref {eq:StwobodyManifold_restructured} captures the nonlinear (two-excitation) scattering process, mapping the incoming  state $|\eta_{\alpha}(E,\bm{P})\rangle$ to the outgoing state $|\bar{\eta}_{\beta}(E,\bm{P})\rangle$ with a dimensionless amplitude $\tau_{\alpha,\beta}(E,\bm{P})$
\begin{equation}
\tau_{\alpha,\beta}(E,\bm{P}) = 2\pi i\, \frac{\sqrt{\rho_{\alpha}(E,\bm{P})\,\rho_{\beta}(E,\bm{P})}}{L(E+i0, \bm{P})}.
\label{eqtauEP}
\end{equation}
\label{eqSonshellTau}
\end{subequations}
The density of states factors in Eq.~\eqref{eqtauEP} naturally arise from projecting the S-matrix onto the energy-momentum manifold.

It is straightforward to show that applying the linear scattering operator to $|\eta_{\alpha}(E,\bm{P})\rangle$ and using the identity $t(k^{(2,\alpha)}) a^*(k^{(2,\alpha)}) = a(k^{(2,\alpha)})$, we have 
\begin{align}
S_{\mathrm{linear},\alpha}^{(2)}(E,\bm{P}) \, |\eta_{\alpha}(E,\bm{P})\rangle 
= |\bar{\eta}_{\alpha}(E,\bm{P})\rangle. \label{eqSlineareta}
\end{align}

Combining Eq.~\eqref{eqSlineareta} with Eq.~\eqref{eqSonshellTau} leads directly to Eq.~\eqref{eqS_on_shell_simple}.

\bibliographystyle{apsrev4-2}
\bibliography{library_corrected}

\appendix
\end{document}